\documentclass[aps,pra,twocolumn,superscriptaddress,floatfix, 10pt, showpacs,final]{revtex4}
\usepackage[usenames]{color}
\usepackage{amsmath}
\usepackage{amssymb}
\usepackage{mathrsfs}
\usepackage{euscript}
\usepackage{dcolumn}
\usepackage{graphicx}
\usepackage{color}

\begin{document}
\title{Indirect Dissociative Recombination of LiH$_2^+$ + $e^-$}
\author{Daniel J. Haxton}
\email[]{dhaxton@jila.colorado.edu}
\affiliation{Department of Physics and JILA, University of Colorado, Boulder Colorado 80309}
\author{Chris H. Greene}
\email[]{chris.greene@colorado.edu}
\affiliation{Department of Physics and JILA, University of Colorado, Boulder Colorado 80309}

\pacs{03.65.Nk, 34.80.-i, 34.80.Lx, 33.20.Wr}

\begin{abstract}

We present the results of calculations determining the cross sections for 
indirect dissociative recombination of LiH$_2^+$ + $e^-$.  These calculations employ 
multichannel quantum defect theory and Fano's
rovibrational frame transformation technique to obtain the indirect DR cross 
section in the manner described by Ref.\cite{hamilton}.  
We use \textit{ab initio} electron-molecule scattering codes to calculate quantum
defects.  In contrast to H$_3^+$, the LiH$_2^+$ molecule exhibits 
considerable mixing between rotation and vibration; however, by incorporating
an exact treatment of the rovibrational dynamics of the LiH$_2^+$, we show that this
mixing has only a small effect on the observed DR rate.  We calculate a large DR rate
for this cation, 4.0 $\times$ 10$^{-7}$ cm$^{3}$ s$^{-1}$ at 1~meV incident electron
energy.

\end{abstract}

\maketitle

\section{Introduction}

Dissociative recombination, the process by which a cation recombines with 
a free electron and dissociates, 
\begin{equation}
AB^+  +  e^- \longrightarrow A + B \ ,
\end{equation}
has received much theoretical and experimental interest in the past two
decades \cite{mitchell,orelbook}.  Innovations at both sides of the scientific process have spurred
this interest.  The development of storage ring experiments\cite{larssonreview} has been the
key innovation on the experimental side.  Storage rings allow the
preparation of cation species that are rovibrationally cold, such that 
a small number of initial rovibrational states are populated.  Such devices
also enable the synchronization of cation and electron beams, such that 
the relative kinetic energy between the two can be precisely controlled.
As a result, DR rate coefficients can be determined with unprecedented
resolution, and structures in the rate coefficient as a function of 
relative kinetic energy may be elucidated.

The current theoretical understanding of the dissociative recombination
process provides two mechanisms by which it may occur.  These mechanisms
are labeled the ``direct'' and the ``indirect'' process.  The direct process
involves temporary capture of the electron into 
a metastable electronic state of the neutral.  Such resonant electron capture 
was pointed out by Bates in 1950\cite{bates}, and quantitatively formulated later
by O'Malley in 1966\cite{Omalley};
it is particularly effective in capturing low-energy
(thermal) electrons when the Born-Oppenheimer
potential energy curve of the metastable neutral state crosses the 
curve of the ground state cation species within the Franck-Condon region of
the latter.  It may also be the only viable mechanism of dissociative
recombination at high incident electron energy.

When there is no Born-Oppenheimer curve of the neutral that crosses
within the Franck-Condon region of the cation, it is the indirect mechanism\cite{indirect}
that is responsible for any observed dissociative recombination.  The indirect
mechanism is favored by low kinetic energy of the electron-cation collision. 
The indirect process, like the direct process, is a resonant
phenomenon; however, in this case the resonances are rovibrational Feshbach resonances,
not electronic resonances as in the direct process.

Until recently, the consistent, accurate mathematical and numerical description of the 
indirect mechanism
was elusive\cite{h3progress}.  Perhaps the most vexing problem was that of the dissociative
recombination of H$_3^+$, because the dissociative recombination of this species
plays an important role in interstellar chemistry, and due to the numerous
failures of theory to accurately predict the  rate observed by experiment.
Adding to the mystery was the considerable spread in experimental results, ranging
from 2.3 $\times$ 10$^{-7}$ to less than 10$^{-10}$ cm$^{-3}$ s$^{-1}$ at 300$^\circ$K\cite{larssonreview}.

However, the theory outlined in
Ref.\cite{hamilton}, involving a frame transformation with Siegert
states representing the outgoing dissociative flux, has been applied to several systems and
has thus far shown consistently good results in predicting indirect DR rates.
A series of theoretical works\cite{kokoo,kokoogreene,
hd2dr, santos} on the
DR of H$_3^+$ and isotopomers
obtained unprecedented agreement with experiment for this difficult system, 
matching both the overall magnitude and
most of the structure of the experimental cross section\cite{h3expt,mccall,tsr}.
Further use of the method has included a study of LiH$^+$\cite{roman, roman2}
that reproduced the experiment of S. Krohn \textit{et al.}~\cite{krohn, krohnprl} extremely well in all but the
lowest part of the measured incident energy range.

In the present article, we examine the dissociative recombination of another species,
namely LiH$_2^+$.  Despite any superficial similarity
to H$_3^+$, the two cations are in fact quite different, and we view these calculations
as a further step toward validating and generalizing the theory.  In 
particular, the rovibrational structure of the LiH$_2^+$ cation is more complicated
than that of H$_3^+$; whereas Ref.\cite{kokoo}, 
and later, Ref.\cite{santos}
obtained excellent agreement with experiment by using a rigid rotor approximation for
the vibrational states of H$_3^+$, the LiH$_2^+$ cation is well described as a Li$^+$ cation
weakly bound to an H$_2$ molecule, which fragments may rotate relatively independently.
Thus, in the present work we incorporate an exact treatment of the rovibrational
Hamiltonian and compare it to a rigid-rotor treatment.  
This work represents the first such exact treatment of the ionic rovibrational
motion for indirect dissociative recombination in a polyatomic species.

This paper is organized as follows.
We briefly introduce the electronic structure of LiH$_2^+$ and LiH$_2$ in Section
\ref{elecsect}.  We use the
the Swedish-Molecule and UK R-matrix\cite{ukrmatrix} codes to calculate fixed-nuclei electron scattering
quantum defect matrices, and we describe these calculations and present the results
in Section \ref{rmatsect}.  
A description of the calculation of 
the rovibrational states of the cation, including an explanation of the coordinate 
system we use, comprises Section \ref{statessect}.  In Section \ref{ecssect} we describe
how we account for the outgoing dissociative flux; we employ a method different from 
that used in previous calculations, using exterior complex scaling\cite{ecs} instead of Siegert
states to enforce outgoing wave boundary conditions on the vibrational basis.  In
Section \ref{transsect} we describe the rovibrational frame transformation and explain
how the nuclear statistics are taken into account.  
Finally, in Section \ref{resultssect} we present the calculated cross
sections.

\section{Electronic structure of LiH$_2^+$ and LiH$_2$ \label{elecsect}}

The ground electronic state of the LiH$_2$ molecule and the LiH$_2^+$ cation are well described 
qualitatively as a Li atom or Li$^+$ cation 
weakly bound to an H$_2$ molecule.  Both states have an equilibrium geometry with equal Li-H
bond lengths, and in such a geometry the molecule belongs to the
C$_{2v}$ point group.  Using the labels appropriate to C$_{2v}$ symmetry, the electronic 
configuration of the cation is 1$a_1^2$ 2$a_1^2$, for overall $^2$A$_1$ symmetry.  The additional
electron for LiH$_2$ goes into the 3$a_1$ orbital (approximately the Li 2$s$ orbital).  
When the Li-H bond lengths are unequal, the molecule belongs to the C$_s$ point group and the 
cation configuration is labeled 1$a'^2$ 2$a'^2$.

Prior calculations\cite{lester1, lester2, ksh, lih21978, wu, dgk, sn, dunne}
have established that the equilibrium geometry of the cation has
$r_{HH}$ = 1.42$a_0$ and $R$ = 3.62$a_0$, where $R$ is the distance between the Li and the
H$_2$ center of mass.  The two body asymptote Li$^+$ + H$_2$ lies only 0.286eV higher\cite{lih2surf}.
The three body asymptote (Li$^+$ + H + H) lies much higher, 5.034eV~\cite{lih2surf}.  The
LiH+ complex is weakly bound with a dissociation energy of 0.112eV and therefore this two-body
breakup channel is essentially isoenergetic with the three-body channel.

Because the excitation energy of the Li$^+$ cation is very high -- 60.92eV -- the lowest-lying
electronic excitations of LiH$_2^+$ correspond to states of the Li atom bound to a H$_2^+$
molecule.  The ionization energies of Li and H$_2$ are 5.39 and 15.43eV, respectively, and
so we expect the first excited state to occur at roughly 10eV.

\section{Fixed nuclei scattering calculations on $e^-$ + LiH$_2^+$ \label{rmatsect}}

\begin{figure}
\begin{center}
\begin{tabular}{c}
\resizebox{0.95\columnwidth}{!}{\includegraphics*[0.7in,1.1in][5.5in,4.2in]{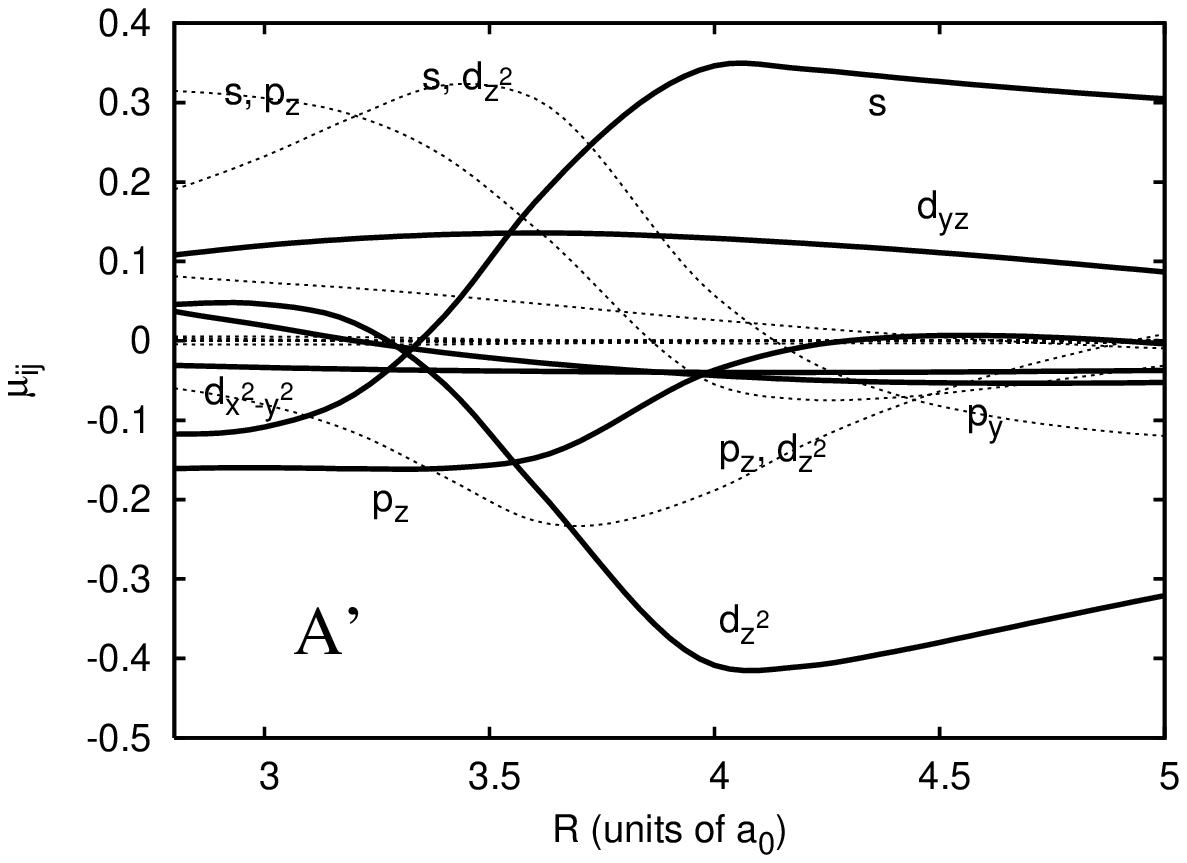}} \\
\resizebox{0.95\columnwidth}{!}{\includegraphics*[0.8in,1.1in][5.5in,2.7in]{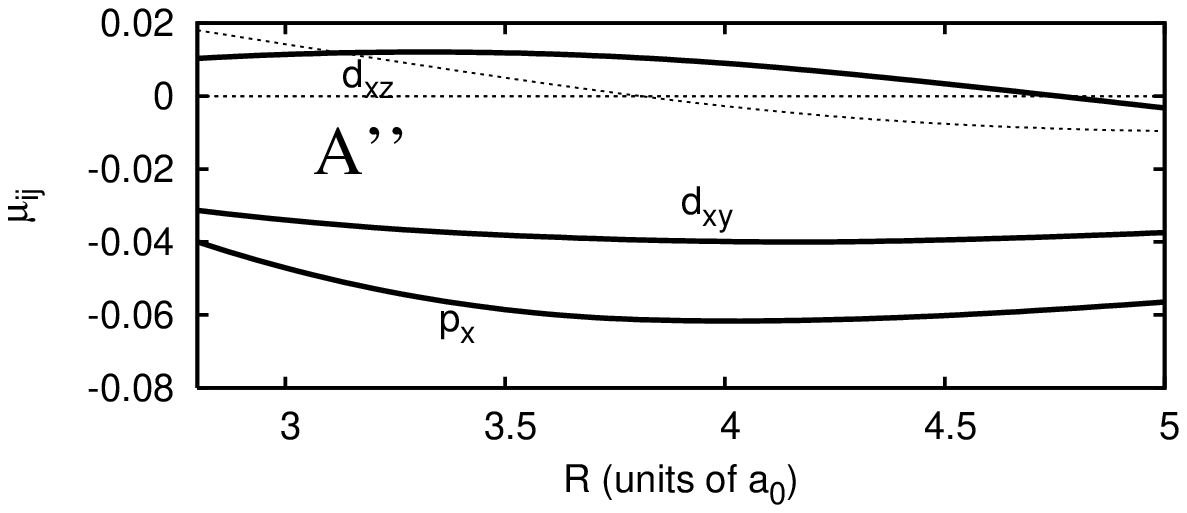}} \\
\resizebox{0.95\columnwidth}{!}{\includegraphics*[0.7in,0.6in][5.5in,2.0in]{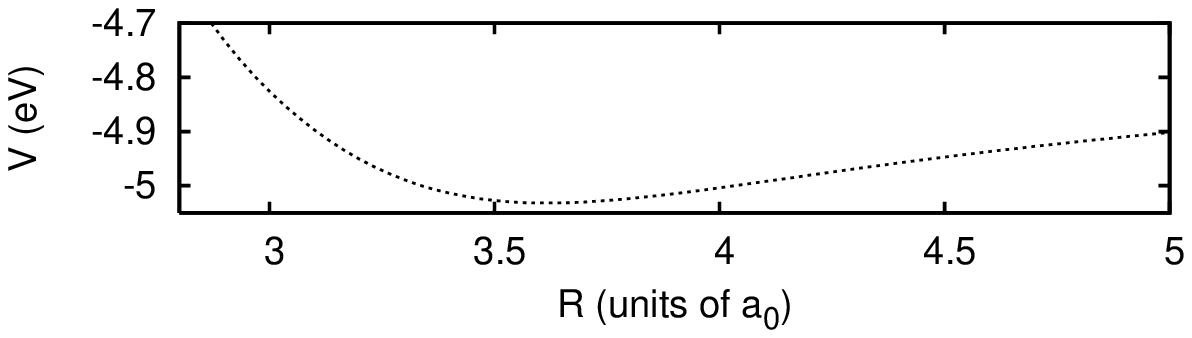}} 
\end{tabular}
\end{center}
\caption{Quantum defect matrix elements (top two panels) and cation potential energy surface\cite{lih2surf} (bottom panel)
as a function of the Jacobi coordinate $R$, fixing $r_{HH}$ = 1.4$a_0$
and $\gamma$ = 90$^\circ$.  Diagonal matrix elements are labeled with a single partial wave; coupling matrix
elements are labeled with the two.  \label{rdefects}}
\end{figure}

\begin{figure}
\begin{center}
\begin{tabular}{c}
\resizebox{0.95\columnwidth}{!}{\includegraphics*[0.7in,1.1in][5.5in,4.2in]{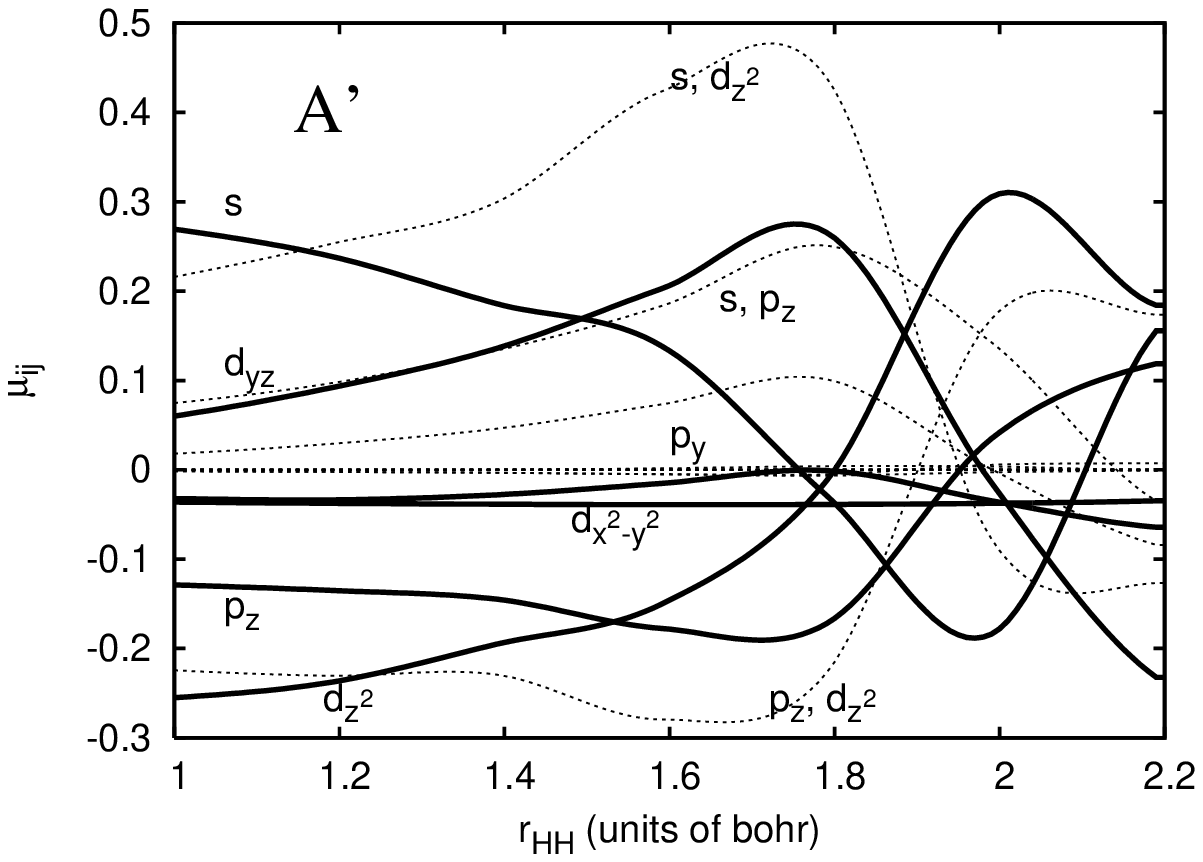}} \\
\resizebox{0.95\columnwidth}{!}{\includegraphics*[0.8in,1.1in][5.5in,2.7in]{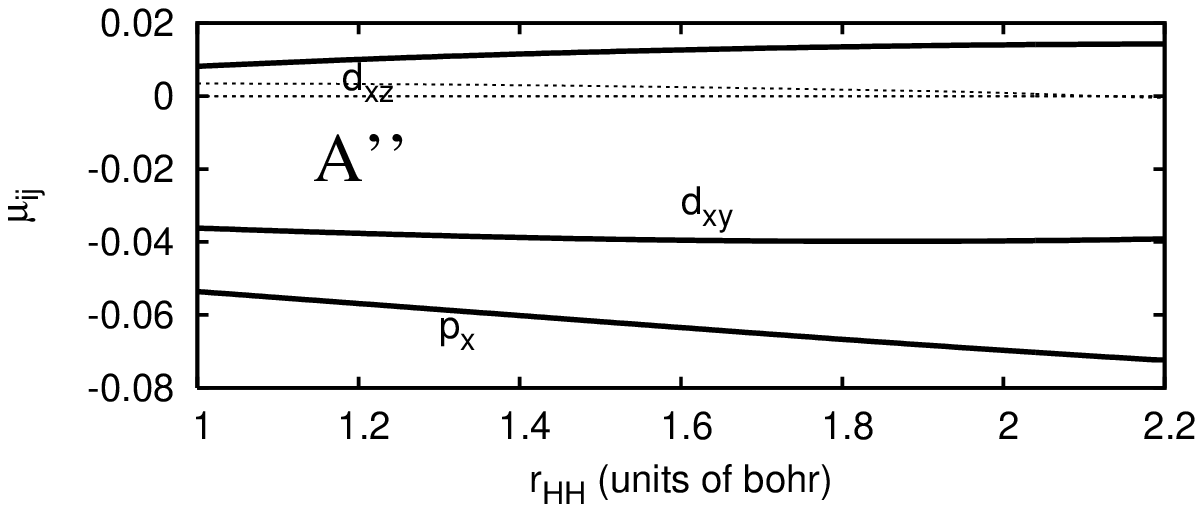}} \\
\resizebox{0.95\columnwidth}{!}{\includegraphics*[0.7in,0.6in][5.5in,2.0in]{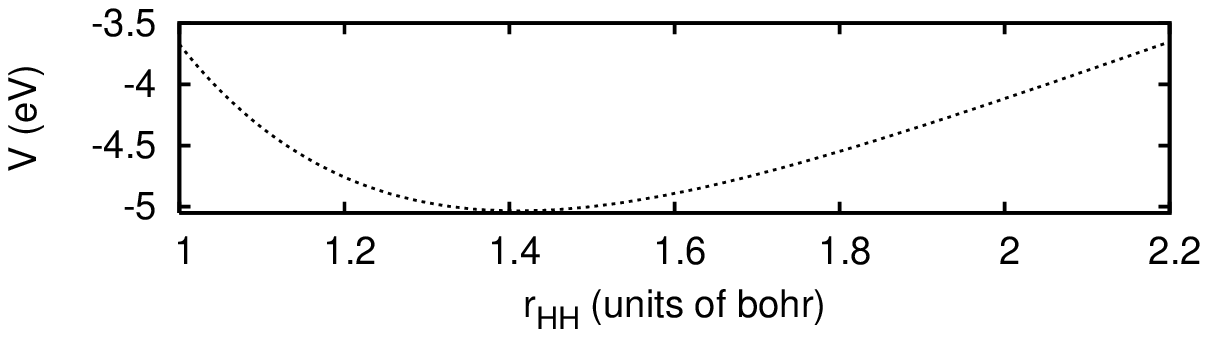}} 
\end{tabular}
\end{center}
\caption{Quantum defect matrix elements  (top two panels) and cation potential energy surface\cite{lih2surf} (bottom panel) as a function of the Jacobi coordinate $r_{HH}$, fixing $R$ = 3.62$a_0$
and $\gamma$ = 90$^\circ$. \label{rhhdefects}}
\end{figure}

\begin{figure}
\begin{center}
\begin{tabular}{c}
\resizebox{0.95\columnwidth}{!}{\includegraphics*[0.7in,1.1in][5.5in,4.2in]{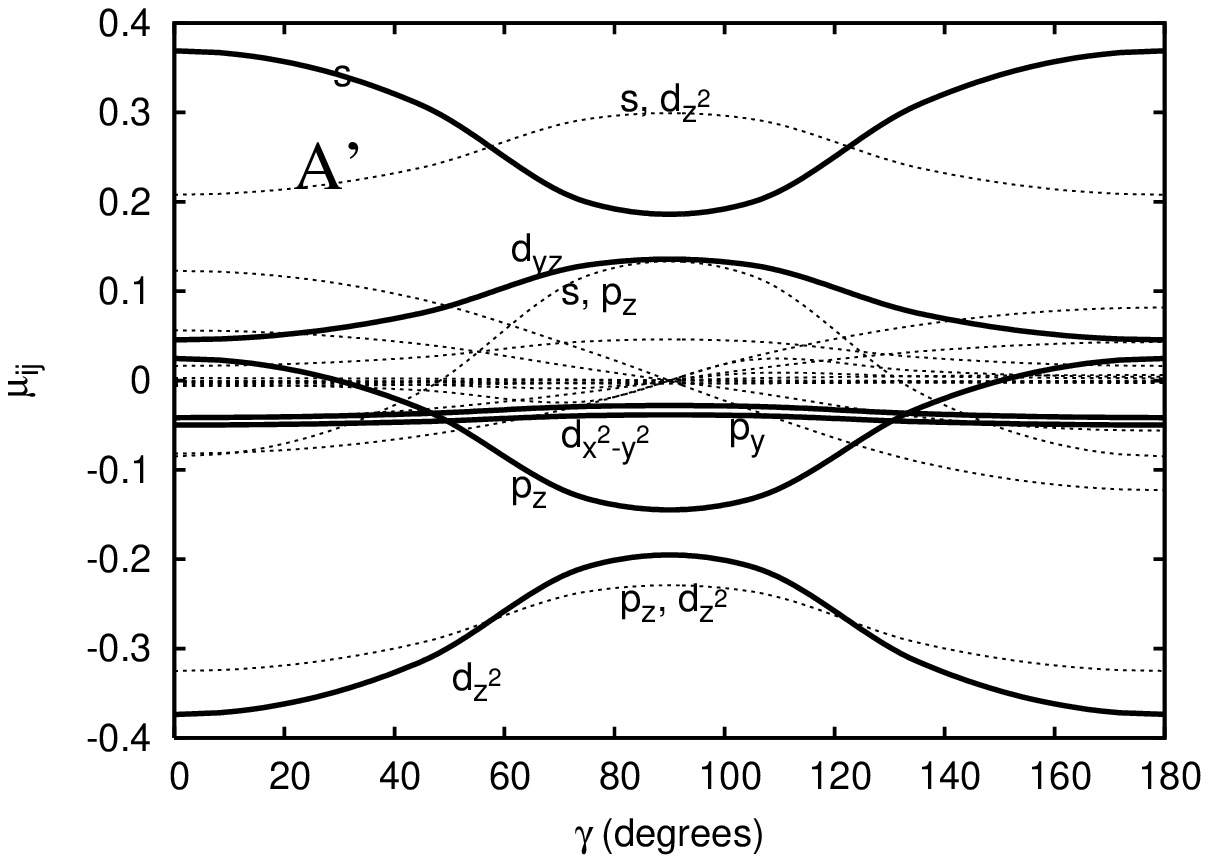}} \\
\resizebox{0.95\columnwidth}{!}{\includegraphics*[0.8in,1.1in][5.5in,2.7in]{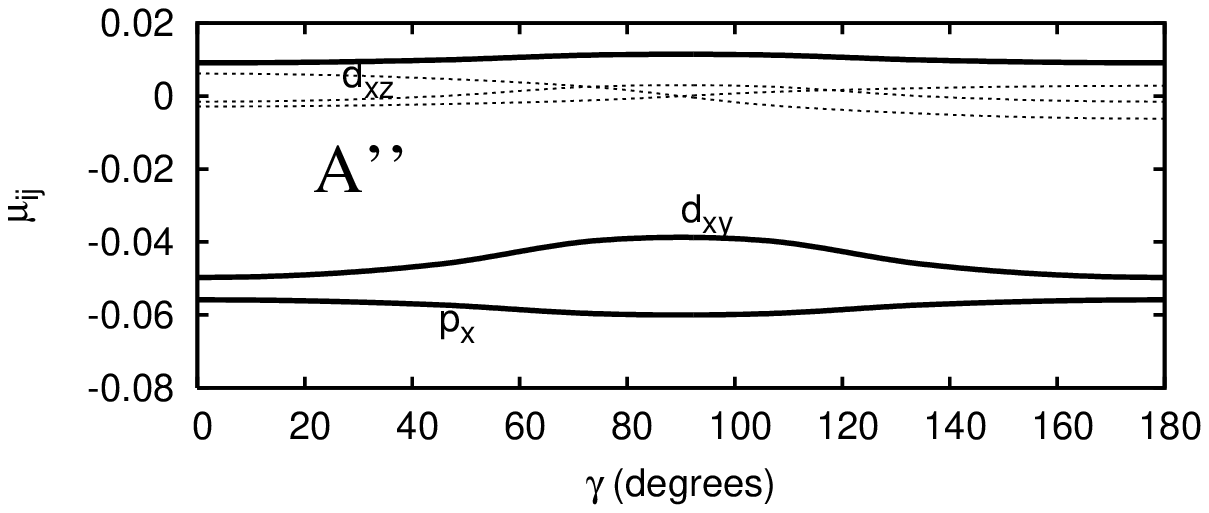}} \\
\resizebox{0.95\columnwidth}{!}{\includegraphics*[0.7in,0.6in][5.5in,2.0in]{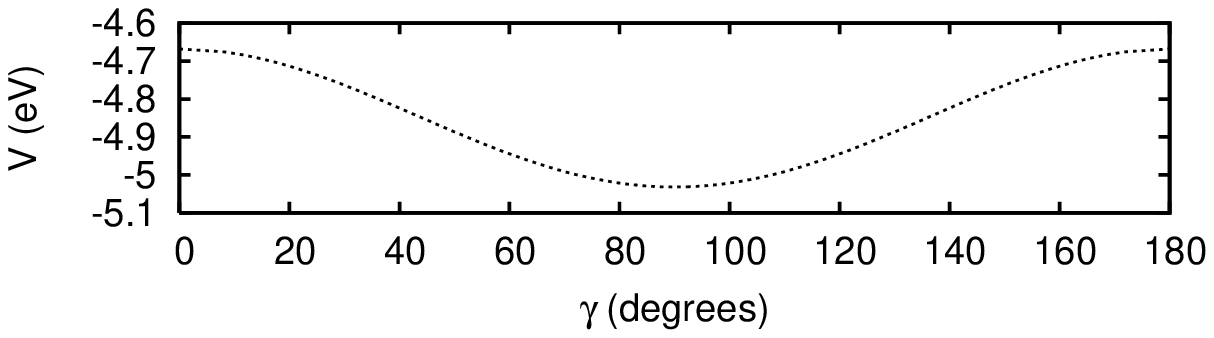}} 
\end{tabular}
\end{center}
\caption{Quantum defect matrix elements  (top two panels) and cation potential energy surface\cite{lih2surf} (bottom panel) 
as a function of the Jacobi coordinate $\gamma$, fixing $R$ = 3.62$a_0$
and $r_{HH}$ = 1.4$a_0$. \label{angdefects}}
\end{figure}

The first step in the present treatment is the calculation of the fixed-nuclei quantum defect
matrices, in the body frame, which describe the scattering of an electron from the LiH$_2^+$ 
cation, with the positions of the nuclei frozen in space.  To perform this task we employ
the polyatomic UK R-matrix scattering codes\cite{ukrmatrix} based on the Swedish-Molecule 
electronic structure suite.

The R-matrix calculation is defined as follows.  We employed an augVTZ STO basis set\cite{lih2basis}
and a 20 bohr spherical R-matrix box radius.  The center of mass of the LiH$_2^+$ cation was placed at the
origin.  We first perform a Hartree-Fock
calculation on the cation using the 1$a'^2$ 2$a'^2$ configuration.  The target wavefunctions are defined as 
having the 1a$'$ orbital (the Li 1s orbital) frozen in double occupation, with the remaining two
electrons distributed among the space 2-6$a'$ and 1a$''$.  We keep the first nine roots of this complete active space configuration-interaction (CAS-CI) calculation to include in the scattering calculation.  These
correspond to the ground state, and excited states that correspond roughly to an H$_2^+$ molecule
bound to a Li atom in its $X$ $^2S$ or $^2P$ configurations, singlet or triplet coupled.  Thus we
have four $^1$A$'$ states, three $^3$A$'$ states, and one $^1$A$''$ and $^3$A$''$ state.  At
the equilibrium geometry of the cation our treatment places these states between 12.86 and 17.13eV.

To the target orbital space we add a set of uncontracted Gaussians that represent the scattering
electron.  This set is obtained using the UK R-matrix code GTOBAS\cite{gtobas}, which optimizes the set
to best fit a set of coulomb wavefunctions orthonormal over the R-matrix sphere.  We include
15 $s$ orbitals, 13 $p$ orbitals, and 12 $d$ orbitals optimized to fit coulomb wavefunctions up
to 10 hartree.

The five-electron space included in the R-matrix
calculation is defined as follows.  We include a close-coupling expansion corresponding to the
first nine states discussed above 
times scattering orbitals, plus penetration terms in which all five electrons are distributed
among the target orbitals, again keeping the 1$a'$ orbital doubly occupied.  The calculation 
is performed in overall A$'$ or A$''$ symmetry.

These calculations yield the fixed-nuclei quantum defect matrices $\mu_{lm,l'm'}$ that are
included in the later steps of the dissociative recombination calculation.  
The quantum defect matrix is defined in terms of the fixed-nuclei S-matrix
as $\mu = -\frac{i}{2\pi}\ln(S)$.
These quantum defect matrices depend weakly on the incident electron energy; we
evaluate them at an incident electron energy of two meV.
We construct an interpolated quantum defect matrix by splining the calculated
quantum defect matrices over the Jacobi coordinate
range $2.4a_0 < R < 5.6a_0$, $0.6a_0 < r_{HH} < 2.4a_0$, all $\gamma$.

Plots of the splined quantum defect surfaces are shown in Figs.~\ref{rdefects},~\ref{rhhdefects}, 
and~\ref{angdefects}.  These figures show three cuts through the quantum
defect surfaces and the corresponding cuts through the cation potential energy
surface; all three points contain the point ($R$=3.62$a_0$,
$r_{HH}$=1.4$a_0$, $\gamma$=90$^\circ$) in Jacobi coordinates.  The cuts are in the $R$ direction
(Fig.~\ref{rdefects}), the $r_{HH}$ direction (Fig.~\ref{rhhdefects}), and the $\gamma$ direction (Fig.~\ref{angdefects}).
The convention in these figures is that all of the diagonal quantum defects
are labeled and labeled with a single channel index, and some of the off-diagonal defects
are labeled and labeled by the corresponding pair of indices.
The molecule lies in the $yz$ plane and the vector $\vec{R}$,
which connects the Li atom to the H$_2$ center of mass, is collinear with the $z$ axis.  

For the calculation in overall A$''$ symmetry there are three electronic channels included 
in the R-matrix calculation, the $p_x$, $d_{xz}$, and $d_{xy}$.  We find that the quantum defects 
in A$''$ symmetry are relatively small.
For the calculation in overall A$'$ symmetry there are six electronic channels
included in the R-matrix calculation.
The quantum defect matrix elements involving $p_y$ and $d_{x^2-y^2}$ are
small relative to the other four.

\section{Calculation of bound and outgoing wave rovibrational states of the cation \label{statessect}}

\begin{figure}
\begin{center}
\begin{tabular}{c}
\resizebox{0.95\columnwidth}{!}{\includegraphics{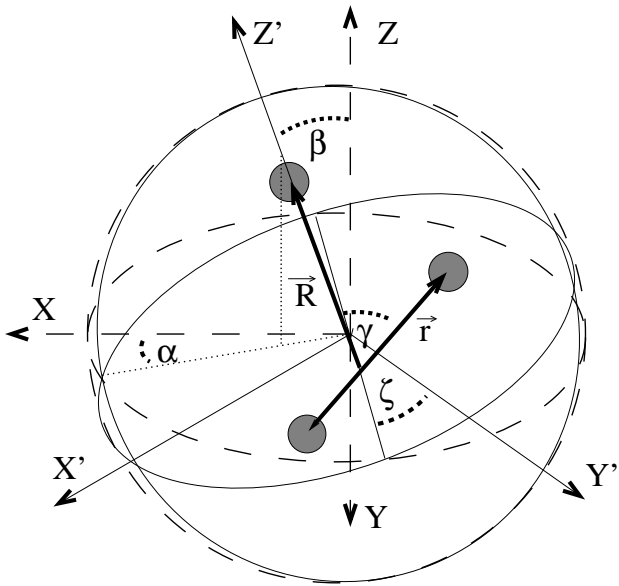}}
\end{tabular}
\end{center}
\caption{``R-embedding'' rovibrational Jacobi coordinate system with origin at the center of mass. 
Primed and unprimed axes refer to BF and SF frames, respectively. The BF $x'z'$ and $x'y'$ planes are both
marked with a thin line circle and the SF $xz$ and $xy$ planes are marked with dashed circles.
The line of nodes is also drawn. The molecule resides in the BF $x'z'$ plane.  The Delves hyperspherical coordinates
$\theta$ and $\mathscr{R}$ used in the DR calculation are defined in terms of $R$ and $r$.
\label{bigcoords}}
\end{figure}

The next step in the DR treatment involves the calculation of rovibrational eigenfunctions
using the ground cation potential energy surface.  We employ the surface of 
Martinazzo \textit{et al.}\cite{lih2surf}, 
which includes the proper long-range behavior of the potential.  

\subsection{Coordinate system and Hamiltonian}

As in previous treatments\cite{kokoo, kokoogreene, hd2, santos}, we use a hyperspherical 
coordinate system and construct
rovibrational states in an adiabatic hyperspherical basis\cite{macek1968}.  The adiabatic expansion
helps to reduce the size of the calculation.

In contrast to the previous treatments we use Delves hyperspherical coordinates\cite{delves1, delves2}.
These coordinates are built from the 
Jacobi coordinate system appropriate to the system, in which $r_{HH}$ denotes the H$_2$ bond
length, $R$ denotes the distance between the Li atom and the H$_2$ center of mass, and $\gamma$
denotes the angle between the two corresponding vectors.  The Delves coordinates consist of 
the Jacobi coordinate $\gamma$, plus two additional coordinates $\mathscr{R}$ and $\theta$,
\begin{equation}
\begin{split}
\mathscr{R} & = \sqrt{R^2 + \frac{\mu_r}{\mu_R} r_{HH}^2} \\
\theta & = \tan^{-1} \left( \sqrt{\frac{\mu_R}{\mu_r}} \frac{R}{r} \right) \ .
\end{split}
\end{equation}

For calculations with nonzero total cation rotational
angular momentum $J^+$, we employ the $R$-embedding coordinate system\cite{rembed} in which the Euler angles
$\alpha, \beta, \zeta$ orient the molecular $z'$ axis, collinear with the $R$ vector, and the
molecular $x'z'$ plane, which contains the molecule, relative to space-fixed axes.  This 
coordinate system is depicted in Figure~\ref{bigcoords}.

We employ the exact rovibrational Hamiltonian for this coordinate system, taken from its form in
Jacobi coordinate system -- see, for example, Refs.\cite{petrongolo,hd2}.
\begin{equation}
\begin{split}
H^{J^+}_{KK}  = & \ \frac{1}{2\mu_\mathscr{R} \mathscr{R}^2} \Big[
 - \frac{\partial^2}{\partial \phi^2} - \frac{1}{4}
+ \frac{1}{\sin^2 \theta \cos^2 \theta} \hat{j}^2 \\
& + \frac{1}{\sin^2 \theta} [J^+(J^++1) - 2K^2 +\hat{j}^2]  \Big] \\
& + V(R,r,\gamma)  - \frac{1}{2\mu_\mathscr{R}} \frac{\partial^2}{\partial \mathscr{R}^2} \\
H^{J^+}_{K \pm 1, K}  =&  \frac{1}{2\mu_\mathscr{R} \mathscr{R}^2 \sin^2 \theta}
 \sqrt{J^+(J^++1) - K(K \pm 1)}  \hat{j}_{\pm} \\
\hat{j}^2  = & \ -\left(\frac{1}{{\sin} 
  (\gamma)}\frac{\partial}{\partial \gamma} {\sin} 
  (\gamma) \frac{\partial}{\partial \gamma} - 
    \frac{K^2}{{\sin}^2 (\gamma)}\right) \\
\hat{j}_{\pm}  = & \ \mp \frac{\partial}{\partial \gamma} - K 
{\mathrm{cot}}(\gamma) \ . 
\label{Hamiltonian}
\end{split}
\end{equation}

In this equation, the operators $\hat{j^2}$ and $\hat{j}_\pm$ are the total and raising/lowering operators of the diatom angular momentum.  This Hamiltonian operates on the expansion coefficients $\chi_K$
in the following expansion of a wavefunction,
\begin{equation}
\Psi_{J^+M} = \sum_{K} \frac{\chi_{K}(\theta,\gamma,\mathscr{R})}{\mathscr{R}^{5/2}} \widetilde{D}^{J^+}_{MK}(\alpha,\beta,\zeta) \ ,
\end{equation}
where the basis of $\widetilde{D}^{J^+}_{MK}(\alpha,\beta,\zeta)$ is the set 
of normalized Wigner rotation matrices (and BF angular momentum eigenstates)
\begin{equation}
\label{expansion}
\widetilde{D}^{J^+}_{MK}(\alpha,\beta,\zeta) = \sqrt{\frac{2J^++1}{8\pi^2}}D^{J^+}_{MK}(\alpha,\beta,\zeta)  \ .
\end{equation}

\subsection{Coupled adiabatic hyperspherical treatment}

The first step in calculating the rovibrational states is to calculate the adiabatic hyperspherical
basis.  Therefore, defining 
$H^{J^+} = H^{J^+}_0(\mathscr{R}) - \mathscr{R}^{-5/2} \frac{1}{2\mu_\mathscr{R}} \frac{\partial^2}{\partial \mathscr{R}^2} \mathscr{R}^{5/2}$ 
where $H^{J^+}_0$ is the adiabatic Hamiltonian, we first solve for 
adiabatic basis functions $\chi^{J^+M}_{j}(\theta,\gamma, \alpha, \beta, \zeta; \mathscr{R})$
and eigenvalues $\epsilon^{J^+}_j(\mathscr{R})$ ,
\begin{equation}
\begin{split}
H^{J^+}_0(\mathscr{R}) & \chi^{J^+M}_{j}(\theta,\gamma, \alpha,\beta, \zeta; \mathscr{R}) \\
& = \epsilon^{J^+}_j(\mathscr{R}) \chi^{J^+M}_{j}(\theta,\gamma, \alpha,\beta,\zeta; \mathscr{R}) \ , \\
\end{split}
\end{equation}
where we expand $\chi^{J^+M}_{j}$ as
\begin{equation}
\chi^{J^+M}_{j}(\theta,\gamma, \alpha,\beta, \zeta; \mathscr{R}) = \sum_{K} \frac{\chi_{jK}^{J^+}(\theta,\gamma; \mathscr{R})}{\mathscr{R}^{5/2}} \widetilde{D}^J_{MK}(\alpha,\beta,\zeta) \ .
\end{equation}

The $\alpha$-th rovibrational eigenfunction for total cation rotational
angular momentum $J^+$ is then expanded as
\begin{equation}
\psi^+_{J^+M\alpha} = \sum_{ij} c^{J^+}_{ij \alpha} \phi_i(\mathscr{R}) \chi^{J^+M}_{j}(\theta,\gamma,\alpha,\beta,\zeta; \mathscr{R}_i) \ .
%
%
\end{equation}
The coefficients $c^{J^+}_{ij \alpha}$ multiply basis functions $\phi_i(\mathscr{R})$ based 
on gridpoints $\mathscr{R}_i$.  These functions comprise a Discrete Variable Representation 
(DVR)\cite{dickcert,lhl,lebdvr}, specifically, the
Gauss-Lobatto finite element DVR\cite{femdvr}
with five elements 1.6 bohr long,
starting at 2.0 bohr, and order 10 within each element.  For the hyperangular degree of freedom $\theta$ 
we also use Gauss-Lobatto DVR, but with one element, and 60th order.  The wavefunction is defined to be
zero at $\theta$ = 0 and 90$^\circ$.  For the $\gamma$ degree of freedom we use 
Legendre DVR based upon associated Legendre functions $P_{lK}$.  The potential is evaluated using
the DVR approximation, which corresponds to a diagonal representation.

\begin{figure}
\begin{center}
\resizebox{1.0\columnwidth}{!}{\includegraphics{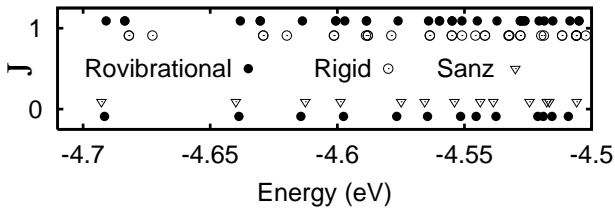}}
\end{center}
\caption{Rovibrational energy eigenvalues for $J^+$=0 and 1 (relative to three-body breakup) 
calculated presently with the surface of Martinazzo \textit{et al.}\cite{lih2surf} and
the full rovibrational Hamiltonian, Eq.(\ref{Hamiltonian}) (filled dots); those calculated for
$J^+$=1 with a rigid rotor approximation (empty dots); and those calculated by Sanz \textit{et
al.}\cite{gianvib} with the Martinazzo surface (triangles).
\label{statesfig}}
\end{figure}

To calculate the full vibrational wavefunctions including the nonadiabatic coupling,
we employ the slow variable discretization of Tolstikhin\cite{svd}, and therefore solve
the matrix equation for the coefficients $c^{J^+}_{ij\alpha}$,
\begin{equation}
\hat{H}^{J^+} \vec{c}^{J^+}_\alpha = E^{J^+}_\alpha \vec{c}^{J^+}_\alpha \ ,
\end{equation}
where the matrix $\hat{H}^{J^+}$ is defined
\begin{equation}
\hat{H}^{J^+}_{ij, i'j'} = \epsilon^{J^+}_j(\mathscr{R}_i) + \hat{O}^{J^+}_{ij, i'j'} \left(T_\mathscr{R}\right)_{ii'} \ ,
\end{equation}
where $T_\mathscr{R}$ is the Gauss-Lobatto kinetic energy matrix for the hyperradius, 
and where the matrix $\hat{O}^{J^+}$ is the overlap matrix 
\begin{equation}
\hat{O}^{J^+}_{ij, i'j'} = \left\langle \chi^{J^+}_{j}(\mathscr{R}_i) \right\vert\left. \chi^{J^+}_{j'}(\mathscr{R}_{i'}) \right\rangle \ ,
\end{equation}
brackets denoting integration over all degrees of freedom except $\mathscr{R}$.

\subsection{Rigid rotor approximation and rovibrational energies}

To calculate the rigid rotor states, we calculate the vibrational states
for total cation angular momentum $J^+ = 0$, obtaining their
wavefunctions $\Psi^+_{00\alpha}$ and energies $E^0_\alpha$.
We find the principal moments of inertia $A$, $B$, and $C$ for each state; 
the largest of these, $A$,
is perpendicular to the molecular plane.  We use this moment as
the axis of quantization and then diagonalize the asymmetric top
hamiltonian
\begin{equation}
\begin{split}
H_{rigid} = & \frac{B+C}{4} \hat{J}^2 + \frac{2A-B-C}{4}\hat{J}_z^2 \\ & + \frac{B-C}{8} \left( \hat{J}^+ \hat{J}^+  +  \hat{J}^- \hat{J}^- \right) \ , 
\end{split}
\end{equation}
in the basis $\widetilde{D}^{J^+}_{MK}$ for a given total cation angular momentum $J^+$.
(In  this equation,
$\hat{J}^\pm$ are raising and lowering operators of the projection, $K$, of the
total angular momentum on the body-fixed axis of quantization.  They are 
not to be confused with the
total cation angular momentum $J^+$, where $J^+(J^++1)$ 
is the eigenvalue
of the total angular momentum squared operator $\hat{J^2}$.  $K$ is the eigenvalue
of $J_z$.  $M$ is, yet again, arbitrary.)
For each value of $J^+$ and each $J^+=0$ state $\Psi^+_{00\alpha}$, 
we obtain $2J^+ + 1$ eigenvalues which are
added to $E^0_\alpha$ to yield the rigid rotor energies for that vibrational
state.  For the purposes of the rotational frame transformation, we transform
the eigenvectors of $H_{rigid}$ such that their axis of quantization,
conjugate to the eigenvalue $K$, is parallel with the Jacobi vector $\vec{R}$,
not perpendicular to the plane.

The $J^+=0$ vibrational energies 
(which are eqiuvalent in the rigid rotor and full rovibrational calculations)
are in
good agreement with the results of Sanz~\textit{et al.}\cite{gianvib}.  
For $J^+ > 0$, The rigid rotor approximation gives significantly different low-lying
eigenvalues than the full rovibrational calculation.
In Figure~\ref{statesfig}
we plot the energies for rovibrational states with $J^+$=0 and 1.  The eigenvalues of Sanz \textit{et al.}
for $J^+$=0 agree reasonably well with ours.  For $J^+$=1 we plot eigenvalues calculated with the full Hamiltonian,
Eq.(\ref{Hamiltonian}), as well as those calculated in the rigid rotor approximation.  One can clearly see
that it is not accurate to treat this molecule as a rigid rotor.

\begin{figure}
\begin{center}
\resizebox{1.0\columnwidth}{!}{\includegraphics{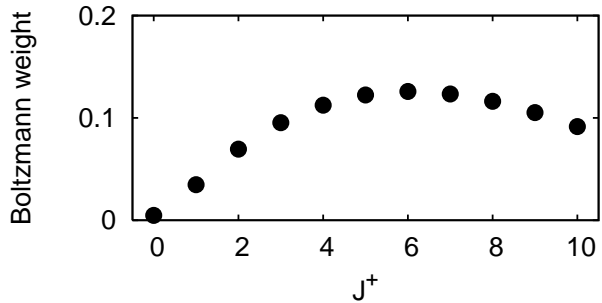}}
\end{center}
\caption{Boltzmann weights at 300$^\circ$ K binned by total cation angular momentum $J^+$.
\label{partition}}
\end{figure}

We plot the Boltzmann weights binned by cation rovibrational angular momentum value $J^+$ in Figure~\ref{partition}.
The number of rovibrational states goes as $(2 J^+ + 1)^2$ and thus the most probable $J^+$ value at 300$^\circ$ K
is six.

\subsection{Nuclear statistics}

The full rovibrational Hamiltonian is invariant with respect to permutations of the two hydrogen
atoms.  Therefore, the rovibrational eigenfunctions will have an eigenvalue of either
+1 or -1 with respect to this permutation operation, which can be expressed 
( $\gamma \rightarrow 90^\circ - \gamma$; $\zeta \rightarrow \zeta + 180^\circ$ ).
Given that the hydrogen atom is a fermion, the +1 states are paired with a singlet
(para) nuclear spin wavefunction, and the -1 states are paired with a triplet (ortho)
nuclear spin wavefunction.  This gives the +1 and -1 states statistical weights of 1 and 3, 
respectively.

The full rotational/rovibrational frame transformation, described later, does not affect the nuclear
statistics.  However, the rovibration-only frame transformation mixes states with different
permutation eigenvalues, and therefore we cannot account for the proper nuclear statistics
with this transformation.

\section{Representation of outgoing flux \label{ecssect}}

The previous implementations of the present theory have employed Siegert pseudostates~\cite{pseudo}
or complex absorbing potentials (CAPs)~\cite{CAPform,CAPref1,CAPref2}
to represent the outgoing flux corresponding to dissociative recombination.  
In contrast, in the current implementation we employ
exterior complex scaling (ECS)\cite{abc1, abc2, moi2, moi3, moi4, moirev, ecs}
to enforce outgoing-wave boundary conditions.
We have found that the use of ECS or CAP states within a MQDT frame-transformation
calculation is more straightforward than the use of Siegert states, as the
completeness and orthogonality 
relationships of the former types of eigenvectors are 
simpler than those of Siegert states.  We will present a more thorough
comparison of these different methods of enforcing outgoing-wave boundary
conditions in a frame transformation calculation in a forthcoming publication.

To calculate the ECS eigenvectors, the final finite element in the $\mathscr{R}$ degree of freedom 
is scaled according to
$\mathscr{R} \rightarrow \mathscr{R}_0 + e^{i\theta}(\mathscr{R}-\mathscr{R}_0)$, where $\mathscr{R}_0$ is the boundary between the fourth
and fifth elements at $\mathscr{R}=8.4a_0$.  We employ a scaling angle of $\frac{1}{8}\pi$.
As with Siegert states, this leads to a discretized 
representation of the dissociative Li$^+$ + H$_2$ vibrational continuum in which the outgoing
wave states have a negative imaginary component to their energy.

Because the coordinate $\mathscr{R}$ is scaled into the complex plane, it is ideal (but often
not necessary\cite{femdvr}) to analytically continue the potential energy surface $V(\mathscr{R},\theta,\gamma)$.
We do so by ensuring that the long-range components to the Martinazzo \textit{et al.} surface 
are evaluated for complex arguments.  We
evaluate their switching formula (third equation on page 11245 of their publication\cite{lih2surf})
by taking the absolute value of the argument.

\section{Rovibrational frame transformation \label{transsect}}

\subsection{Introduction}

The rovibrational frame transformation comprises the central part of the present
calculation.  Frame transformation techniques were originally developed by Fano\cite{fanoft,changfano}  and
have found much use in atomic and molecular theory.  The central idea of a frame transformation
is to take an S-matrix, which is labeled by incoming and outgoing channel indices, and
transform that S-matrix to a new channel basis.  In its simplest incarnation, adopted here, 
this transformation is
exact if the fixed-nuclei quantum defects are constant with respect to energy. 
The transformation is accomplished via a 
unitary matrix that relates the first set of channels to the second.   Usually, the first set 
of channel indices are appropriate
to describe the system when the scattered electron is near the atomic or molecular target,
and the second set of channel indices are appropriate when the electron has escaped far
from the target. The coeffients of the original rotational frame transformation for a diatomic 
molecule\cite{fanoft} are simply Clebsch-Gordan coefficients.  Other unitary transformations
may be applied for different physical situations: for the calculation of Stark states\cite{starkft},
to transform between $LS$ and $JJ$ coupling\cite{lsjj0, lsjj}, or to transform between molecular 
Hund's cases\cite{jungenraseev}.

The frame transformation is applied to molecular vibration in 
much the same way it is applied to rotation.  When the scattered electron is close to the
molecule, it is moving very fast compared to the molecular framework, and therefore the
scattering may be calculated by fixing the nuclei and obtaining fixed-nuclei, body-frame 
S-matrices $s_{lm, l'm'}(\vec{q})$ where $\vec{q}$ are the internal coordinates of the 
molecule and $lm, l'm'$ label the partial wave electron scattering channels in the body
frame.  The frame transformation provides that the full S-matrix, which has vibrational
channel indices as well as electronic channel indices, is found via 
$S_{lm\alpha, l'm'\beta} = \langle \chi_\alpha \vert s_{lm, l'm'} \vert \chi_\beta \rangle$
where the brackets denote integration over the internal degrees of freedom $\vec{q}$.

It is important to note that this vibrational frame transformation is different from the
Chase approximation\cite{chase}.
The frame transformation is applied to the ``short-range'' S-matrices of Multichannel
Quantum Defect Theory (MQDT)\cite{seaton,general,general2}, which have indices including not only 
open but also closed channels.  As a result, complicated nonadiabatic effects caused
by the long-range potential (here a coulomb potential) may be 
accounted for by the theory\cite{rossh2_1, rossh2_2}.

The most accurate versions of the vibrational frame transformation theory\cite{gaogreene, gaogreene2, gaoh2}
incorporate the energy dependence of the fixed-nuclei S-matrix.  We do not do so and
instead evaluate the fixed-nuclei S-matrices at 2meV, implicitly making the assumption that
these S-matrices are constant with respect to incident electron energy.

We note that many other treatments of dissociative recombination within MQDT have been
devised.  See, for example, Refs.\cite{unified, leemqdt}.  However, the current formulation
is perhaps the most easily applicable to a polyatomic molecule.

\subsection{Rovibrational frame transformations for the asymmetric top}

Child and Jungen\cite{childjungen} have already derived the rotational frame
transformation for the asymmetric top.  We perform both a rovibration-only
frame transformation and a rovibrational/rotational frame transformation that uses
their result.

For the vibration-only frame transformation we calculate
\begin{equation}
S^{J^+}_{\alpha lm, \beta l'm'} =  \left\langle \psi^+_{J^+M\alpha}  \right\vert s_{lm,l'm'}  \left\vert \psi^+_{J^+ M \beta} \right\rangle \ ,
\end{equation}
where value of the index $M$ is irrelevant.

The rovibrational frame transformation of Child and Jungen\cite{childjungen} will not be repeated
in full detail here.  It comprises a square unitary transformation matrix for each value of $J$ (total angular momentum)
and $l$ (the angular momentum of the electron).  It transforms from the body-fixed representation,
with quantum numbers $m$ and $K$ -- denoting the projection of the electron angular momentum about
the molecular axis and the projection of total angular momentum -- to the space-fixed representation, 
with quantum numbers $J^+$ and $K^+$, denoting
the total angular momentum of the cation and its projection.  The body-fixed S-matrices are independent
of $K$.  Thus, 
\begin{widetext}
\begin{equation}
s^J_{l J^+ K^+, l' {J^+}' {K^+}'}(\mathscr{R},\theta,\gamma) = \sum_{m K m' } U^{Jl}_{m K, J^+ K^+} s_{lm, l'm'}(\mathscr{R},\theta,\gamma) U^{Jl'}_{m' K, {J^+}' {K^+}'} \ .
\end{equation}
The full rovibrationally and rotationally transformed S-matrix is then
\begin{equation}
S^{J}_{J^+ \alpha l, {J^+}' \beta l'} = \left\langle \psi^+_{J^+M\alpha}  \right\vert
 \Bigg[ \sum_{K^+ {K^+}'} \left\vert K^+ \right\rangle s^J_{l J^+ K^+,l' {J^+}' {K^+}'} \left\langle {K^+}' \right\vert \Bigg] \left\vert \psi^+_{J^+ M \beta} (\mathscr{R}_i) \right\rangle \ .
\end{equation}
The index $M$ is again irrelevant.
\end{widetext}

\subsection{Channel closing and dissociative recombination cross section}

The final step in the present theory is the construction of the physical, open-channel
S-matrix in terms of the closed-channel S-matrices calculated from the frame transformation.
Whereas the latter are assumed to be energy-independent, a strong energy dependence is introduced
to the former by the formula\cite{kokoo}
\begin{equation}
\begin{split}
\mathscr{S}(E) & = S_{oo} - S_{oc}\left( S_{cc} - e^{-2i\beta(E)} \right)^{-1} S_{co} \\
\beta(E) & = \frac{\pi \delta_{ij}}{\sqrt{2(E_i-E)}} \ , \\
\end{split}
\end{equation}
where the subscript $c$ and $o$ denote the closed and open channel subblocks of the
MQDT S-matrix  $S^{J^+}$ or $S^J$, and we introduce the notation $\mathscr{S}$ for the physical S-matrix.

Because the higher-energy rovibrational states lie above the dissociation energy to Li$^+$ + H$_2$, 
they have outgoing-wave components and negative imaginary components to their energy.  As a result,
the physical S-matrix is subunitary and we assign the missing part to dissociative recombination.
Thus, for the vibration-only transform, we sum over the contributions of each partial wave in the
electronic channel,
\begin{equation}
\sigma^{J^+}_\alpha(E) = \frac{\pi}{2E} \sum_{lm} \left( 1- \sum_{l'm'\beta} \left\vert \mathscr{S}^{J+}_{lm\alpha, l'm'\beta} \right\vert^2 \right) \ ,
\end{equation}
and for the full rotational plus vibrational frame transformation,
\begin{equation}
\sigma^J_{J^+\alpha}(E) = \frac{\pi}{2E} \sum_{l} \left( 1- \sum_{{J^+}'l'\beta} \left\vert \mathscr{S}^{J}_{J^+l\alpha, {J^+}'l'\beta} \right\vert^2 \right) \ .
\end{equation}
where $\alpha$ and $J^+$ denote the initial rovibrational state.

We Boltzmann-average these results, assuming a cation temperature of 300$^\circ$ K.  Thus~\cite{kokoo},
\begin{equation}
\sigma_{vib}(E) = \frac{1}{\Xi} \sum_{J^+ \alpha} \ (2J^+ +1) \ \sigma^{J^+}_{\alpha}(E) e^{\frac{-E_{J^+ \alpha}}{kT}}  
\end{equation}
\begin{equation}
\sigma_{rot}(E) = \frac{1}{\Xi} \sum_{J J^+ \alpha} \ \frac{2J+1}{2J^+ +1} \ \sigma^J_{J^+ \alpha}(E) e^{\frac{-E_{J^+ \alpha}}{kT}}  
\end{equation}
\begin{equation}
\Xi = \sum_{J^+ \alpha} \ (2J^+ +1) \ e^{\frac{-E_{J^+ \alpha}}{kT}}
\end{equation}
with T=300$^\circ$ K.

Finally, we convolute the results with respect to the uncertainty in the incident electron
kinetic energy.  For the present results we use a standard deviation of $\sqrt{2}$~meV in both
the parallel and transverse directions, and perform the averaging as described in Ref.\cite{roman2}.

\section{Results: dissociative recombination cross sections \label{resultssect}}

\begin{figure}
\begin{center}
\begin{tabular}{c}
\resizebox{1.0\columnwidth}{!}{\includegraphics{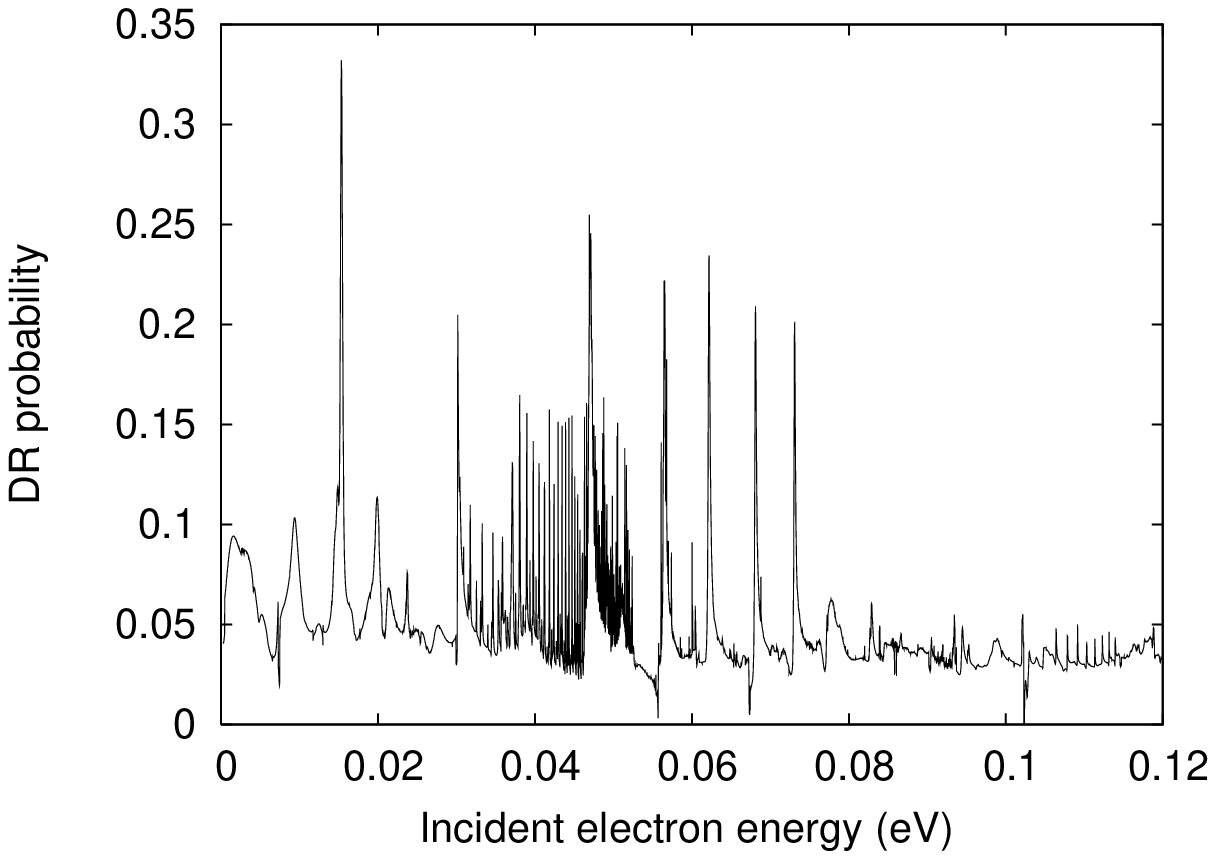}}  \\
\resizebox{1.0\columnwidth}{!}{\includegraphics{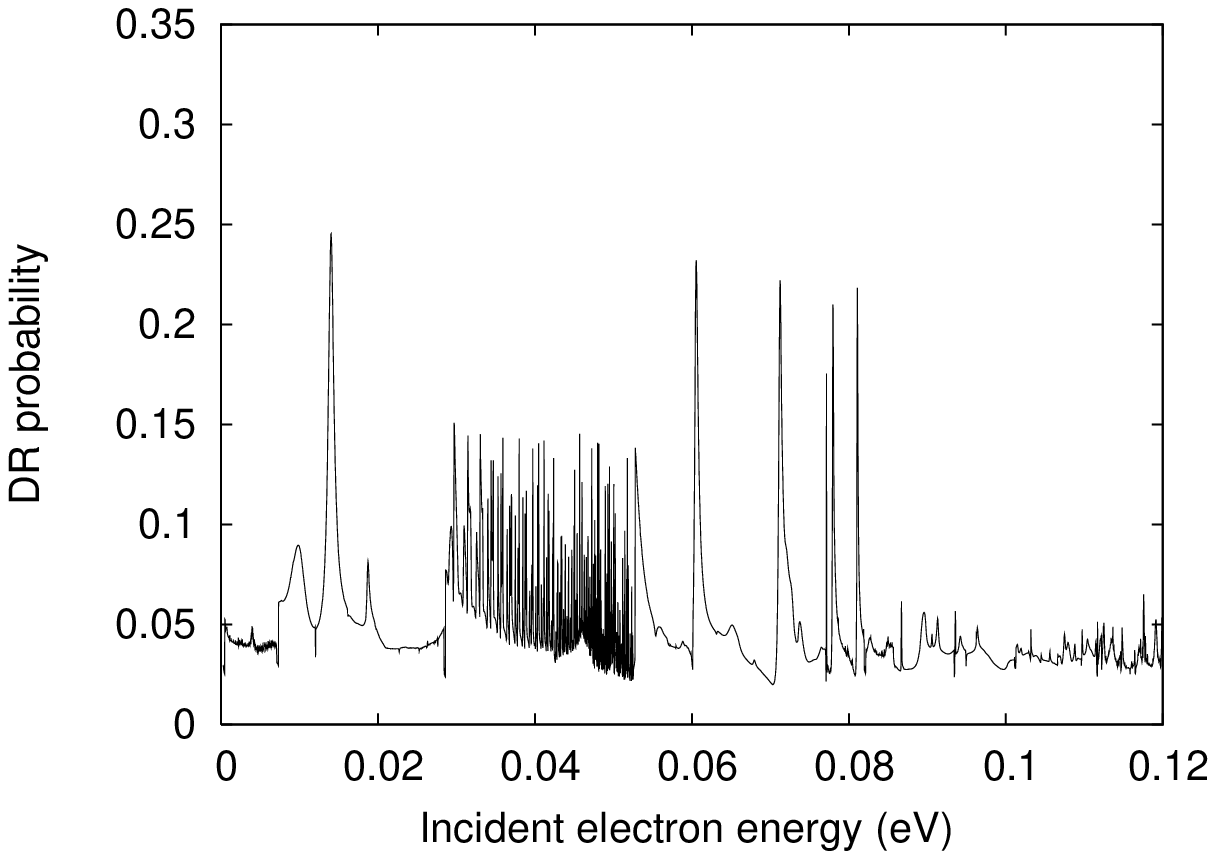}} 
\end{tabular}
\end{center}
\caption{Raw output of frame transformation calculation.  Plotted is the
dissociative recombination probability for the $J^+$=2 vibrational ground state, $s$-wave
channel.  Top, result with full rovibrational Hamiltonian; bottom,
result with rigid rotor states.
\label{rawfig}}
\end{figure}

\begin{figure}
\begin{center}
\begin{tabular}{c}
\resizebox{1.0\columnwidth}{!}{\includegraphics{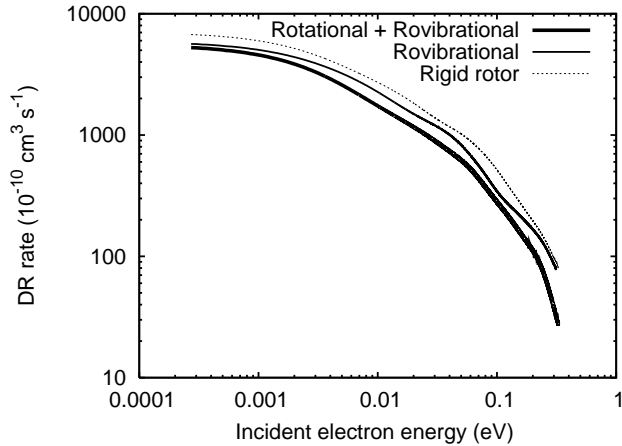}}  \\
\end{tabular}
\end{center}
\caption{Calculated dissociative recombination rate at 300$^\circ$K, assuming experimental resolution of $\sqrt{2}$~meV in the
parallel and transverse directions.  The rotational plus rovibrational curve is our
final result, and the other two curves are from the rovibrational transform only,
using rigid rotor or full rovibrational states.
\label{ratefig}}
\end{figure}

We seek to determine how relevant the inclusion of the exact cation rovibrational dynamics
is to the experimentally observed DR rate.  The raw DR cross sections that we calculate
show considerable structure that depends upon whether an exact or rigid-rotor treatment of
the rovibrational dynamics is used.  However, experiments operate with a thermal sample of
cation targets, including many rovibrational states, and use a beam of electrons with a 
small spread in energies.
Storage-ring experiments are performed with cool cation targets, with rovibrational
temperatures typically on the order of 300$^\circ$ K.  In order to compare with results
obtained under these conditions, we Boltzmann-average over approximately 300 initial
rovibrational states of the LiH$_2^+$ cation, and account for the uncertainty in the
incident electron energy, taken here to be 2meV ($\sqrt{2}$meV in the parallel and 
transverse directions).  In doing so, much of the structure in the
DR cross section is lost, and we find that the rigid rotor treatment is probably sufficient
for calculating rates to be compared with experiment.

An example of the structure in the unconvolved cross sections is shown in Fig.~\ref{rawfig}.
There we show raw results of the rovibration-only frame transformation calculation for $J^+$=2,
both using the full rovibrational Hamiltonian to calculate the rovibrational states, and using a 
rigid rotor approximation for the rovibrational states.  The results are markedly different, 
showing that the strong mixing of rotation and vibration in 
LiH$_2^+$, even at low $J$, affects the structure in the cross sections for 
individual entrance and exit channels.

The first excited rovibrational state lies at 7.3meV.  The sixth and ninth excited state, corresponding to 
excitation in the dissociative $R$ direction and excitation in the
$\gamma$ direction -- rotation of the H$_2$ -- 
lie at 53meV and 77meV, respectively.  As is clear from Figure \ref{rawfig}, there is a prominent
series of narrow rydberg resonances converging to the 53meV threshold, which
serve to enhance the DR rate.  
It is therefore clear that excitation in the dissociative 
direction plays the largest role in the indirect DR process for this molecule, as opposed to rotational
excitation or excitation in the H$_2$ stretch coordinate.

For the purpose of calculating rates to be compared with experiment, we find that
the rigid rotor treatment is probably sufficient, though it apparently overestimates the
cross section slightly.  Not including the rotational frame transformation of Child and Jungen,
we have compared
the rovibration-only frame transformation using the full
rovibrational states to that using the rigid rotor states.  
We find that the rigid-rotor treatment yields
a DR rate consistently about 20\% higher than the
full rovibrational treatment.  
The full calculation, employing the
rotation/rovibrational frame transformation and the full rovibrational states, was not completed,
due to numerical difficulty.  Instead, we perform the rotational transformation of Child and Jungen
with the rovibrationally transformed S-matrix calculated from rigid-rotor states. 
On the basis of the comparison between the calculations not including the rotational transformation
of Child and Jungen, we estimate that this treatment probably overestimates the cross section by
about 20\%.

Our convolved results are shown in Figure \ref{ratefig}.  
We show the DR rate calculated at 300$^\circ$K and including states up to
$J^+$=9 (for the rovibration-only transformations) or $J$=11 (for the rotational and rovibrational transformation).
We show three results: from using the full rovibrational Hamiltonian, with no 
rotational transform; from using a rigid rotor approximation, with no rotational transform; and 
from using a rigid rotor approximation, with the rotational transformation of Child and Jungen.
The former two calculations demonstrate the effect of including the full rovibrational
dynamics, and as mentioned immediately above, the rigid rotor result exceeds the 
full rovibrational result by approximately 20 percent, which factor
is fairly independent of the incident electron energy.
The latter calculation should be considered our final result, 
with the caveat that it probably overestimates
the rate by about 20\%.
Nuclear statistics are included for the full rotational/rovibrational transformation, but not 
for the rovibration-only transformation, because the rovibration-only frame transformation destroys
the permuation symmetry of the overall wavefunction.
The rates are comperable but a bit higher than the corresponding rates for H$_3^+$, by a factor
of two or three.  
The effect of including the rotational part of the transformation is to further lower
the results by about 10\% in the low-energy region, and 50\% in the high-energy region.

\section{Conclusion}

We have applied the method of Ref.~\cite{hamilton} to the calculation of the
indirect DR rate of LiH$_2^+$ + $e^-$.  A central aim of our treatment was to analyze
the effect of including the full rovibrational dynamics of the cation.  We have found
that although the full rovibrational treatment produces channel energies and
unconvolved cross sections considerably different from a rigid rotor treatment,
a rigid rotor treatment is amenable to the calculation of convolved cross sections
to be compared with experiment, although it probably overestimates the DR rate
for a floppy molecule such as LiH$_2^+$ by a small and energy-independent amount.  

The main approximation in the present treatment is the use of energy-independent
quantum defects.  For the calculation of indirect DR rates for the present system, 
this approximation is
expected to be very good, because the amount of energy transferred from electronic
to nuclear motion is rather small due to the small dissociation energy of LiH$_2^+$.  Methods to accurately treat the energy dependence of the fixed-nuclei
s-matrix within a frame transformation exist~\cite{gaogreene, gaogreene2, gaoh2} and may be applied to this system in future work.

The calculations presented here demonstrate that the indirect mechanism provides
a powerful mechanism for dissociative recombination of LiH$_2^+$ + e$^-$.  Future
work will seek to analyze the branching ratios for two- and three-body dissociation and to
further study the nature of the indirect DR mechanism.

\begin{acknowledgments}

We would like to acknowledge Richard Thomas of Albanova University for stimulating
discussions and for sharing experimental results prior to publication.
We acknowledge support under 
DOE grant number W-31-109-ENG-38 and
NSF grant number ITR 0427376, and acknowledge
the National Energy Research Scientific Computing
Center (NERSC) of the DOE Office of Science, which facility was used to
calculate some of the fixed-nuclei quantum defects used in this study.
We would also like to acknowledge the
various entities at JILA that are responsible for funding JILA's yotta
computer cluster.

\end{acknowledgments}

\bibliography{LiH2.bib}

\begin{thebibliography}{72}
\expandafter\ifx\csname natexlab\endcsname\relax\def\natexlab#1{#1}\fi
\expandafter\ifx\csname bibnamefont\endcsname\relax
  \def\bibnamefont#1{#1}\fi
\expandafter\ifx\csname bibfnamefont\endcsname\relax
  \def\bibfnamefont#1{#1}\fi
\expandafter\ifx\csname citenamefont\endcsname\relax
  \def\citenamefont#1{#1}\fi
\expandafter\ifx\csname url\endcsname\relax
  \def\url#1{\texttt{#1}}\fi
\expandafter\ifx\csname urlprefix\endcsname\relax\def\urlprefix{URL }\fi
\providecommand{\bibinfo}[2]{#2}
\providecommand{\eprint}[2][]{\url{#2}}

\bibitem[{\citenamefont{Hamilton and Greene}(2002)}]{hamilton}
\bibinfo{author}{\bibfnamefont{E.~L.} \bibnamefont{Hamilton}} \bibnamefont{and}
  \bibinfo{author}{\bibfnamefont{C.~H.} \bibnamefont{Greene}},
  \bibinfo{journal}{Phys. Rev. Lett.} \textbf{\bibinfo{volume}{89}},
  \bibinfo{pages}{263003} (\bibinfo{year}{2002}).

\bibitem[{\citenamefont{Florescu-Mitchell and Mitchell}(2006)}]{mitchell}
\bibinfo{author}{\bibfnamefont{A.~I.} \bibnamefont{Florescu-Mitchell}}
  \bibnamefont{and} \bibinfo{author}{\bibfnamefont{J.~B.~A.}
  \bibnamefont{Mitchell}}, \bibinfo{journal}{Physics Reports}
  \textbf{\bibinfo{volume}{430}}, \bibinfo{pages}{277} (\bibinfo{year}{2006}).

\bibitem[{\citenamefont{Larsson and Orel}(2008)}]{orelbook}
\bibinfo{author}{\bibfnamefont{M.}~\bibnamefont{Larsson}} \bibnamefont{and}
  \bibinfo{author}{\bibfnamefont{A.~E.} \bibnamefont{Orel}},
  \emph{\bibinfo{title}{Dissociative Recombination of Molecular Ions}}
  (\bibinfo{publisher}{Cambridge University Pres}, \bibinfo{address}{New York},
  \bibinfo{year}{2008}).

\bibitem[{\citenamefont{Larsson}(1997)}]{larssonreview}
\bibinfo{author}{\bibfnamefont{M.}~\bibnamefont{Larsson}},
  \bibinfo{journal}{Annu. Rev. Phys. Chem.} \textbf{\bibinfo{volume}{48}},
  \bibinfo{pages}{151} (\bibinfo{year}{1997}).

\bibitem[{\citenamefont{Bates}(1950)}]{bates}
\bibinfo{author}{\bibfnamefont{D.~R.} \bibnamefont{Bates}},
  \bibinfo{journal}{Phys. Rev.} \textbf{\bibinfo{volume}{78}},
  \bibinfo{pages}{492} (\bibinfo{year}{1950}).

\bibitem[{\citenamefont{O'Malley}(1966)}]{Omalley}
\bibinfo{author}{\bibfnamefont{T.~F.} \bibnamefont{O'Malley}},
  \bibinfo{journal}{Phys. Rev.} \textbf{\bibinfo{volume}{150}},
  \bibinfo{pages}{14} (\bibinfo{year}{1966}).

\bibitem[{\citenamefont{Guberman}(1994)}]{indirect}
\bibinfo{author}{\bibfnamefont{S.~L.} \bibnamefont{Guberman}},
  \bibinfo{journal}{Phys. Rev. A} \textbf{\bibinfo{volume}{49}},
  \bibinfo{pages}{R4277} (\bibinfo{year}{1994}).

\bibitem[{\citenamefont{Orel et~al.}(2000)\citenamefont{Orel, Schnieder, and
  Suzor-Weiner}}]{h3progress}
\bibinfo{author}{\bibfnamefont{A.~E.} \bibnamefont{Orel}},
  \bibinfo{author}{\bibfnamefont{I.~F.} \bibnamefont{Schnieder}},
  \bibnamefont{and}
  \bibinfo{author}{\bibfnamefont{A.}~\bibnamefont{Suzor-Weiner}},
  \bibinfo{journal}{Philos. Trans. R. Soc. London A}
  \textbf{\bibinfo{volume}{358}}, \bibinfo{pages}{2445} (\bibinfo{year}{2000}).

\bibitem[{\citenamefont{Kokoouline and Greene}(2003)}]{kokoo}
\bibinfo{author}{\bibfnamefont{V.}~\bibnamefont{Kokoouline}} \bibnamefont{and}
  \bibinfo{author}{\bibfnamefont{C.~H.} \bibnamefont{Greene}},
  \bibinfo{journal}{Phys. Rev. A} \textbf{\bibinfo{volume}{68}},
  \bibinfo{pages}{012703} (\bibinfo{year}{2003}).

\bibitem[{\citenamefont{Kokoouline and Greene}(2004)}]{kokoogreene}
\bibinfo{author}{\bibfnamefont{V.}~\bibnamefont{Kokoouline}} \bibnamefont{and}
  \bibinfo{author}{\bibfnamefont{C.~H.} \bibnamefont{Greene}},
  \bibinfo{journal}{Faraday Discuss.} \textbf{\bibinfo{volume}{127}},
  \bibinfo{pages}{413} (\bibinfo{year}{2004}).

\bibitem[{\citenamefont{Kokoouline and Greene}(2005)}]{hd2dr}
\bibinfo{author}{\bibfnamefont{V.}~\bibnamefont{Kokoouline}} \bibnamefont{and}
  \bibinfo{author}{\bibfnamefont{C.~H.} \bibnamefont{Greene}},
  \bibinfo{journal}{Phys. Rev. A} \textbf{\bibinfo{volume}{72}},
  \bibinfo{pages}{022712} (\bibinfo{year}{2005}).

\bibitem[{\citenamefont{dos Santos et~al.}(2007)\citenamefont{dos Santos,
  Kokoouline, and Greene}}]{santos}
\bibinfo{author}{\bibfnamefont{S.~F.} \bibnamefont{dos Santos}},
  \bibinfo{author}{\bibfnamefont{V.}~\bibnamefont{Kokoouline}},
  \bibnamefont{and} \bibinfo{author}{\bibfnamefont{C.~H.}
  \bibnamefont{Greene}}, \bibinfo{journal}{J. Chem. Phys.}
  \textbf{\bibinfo{volume}{127}}, \bibinfo{pages}{124309}
  (\bibinfo{year}{2007}).

\bibitem[{\citenamefont{Glosik et~al.}(2001)\citenamefont{Glosik, Plasil,
  Poterya, Kurdna, Rusz, Tichy, and Pysanenko}}]{h3expt}
\bibinfo{author}{\bibfnamefont{J.}~\bibnamefont{Glosik}},
  \bibinfo{author}{\bibfnamefont{R.}~\bibnamefont{Plasil}},
  \bibinfo{author}{\bibfnamefont{V.}~\bibnamefont{Poterya}},
  \bibinfo{author}{\bibfnamefont{P.}~\bibnamefont{Kurdna}},
  \bibinfo{author}{\bibfnamefont{J.}~\bibnamefont{Rusz}},
  \bibinfo{author}{\bibfnamefont{M.}~\bibnamefont{Tichy}}, \bibnamefont{and}
  \bibinfo{author}{\bibfnamefont{A.}~\bibnamefont{Pysanenko}},
  \bibinfo{journal}{J. Phys. B} \textbf{\bibinfo{volume}{34}},
  \bibinfo{pages}{L485} (\bibinfo{year}{2001}).

\bibitem[{\citenamefont{McCall et~al.}(2003)\citenamefont{McCall, Honeycutt,
  Saykally, Geballe, Djuric, Dunn, Semaniak, Novotny, Al-Khalili, Ehlerding
  et~al.}}]{mccall}
\bibinfo{author}{\bibfnamefont{B.~J.} \bibnamefont{McCall}},
  \bibinfo{author}{\bibfnamefont{A.~J.} \bibnamefont{Honeycutt}},
  \bibinfo{author}{\bibfnamefont{R.~J.} \bibnamefont{Saykally}},
  \bibinfo{author}{\bibfnamefont{T.~R.} \bibnamefont{Geballe}},
  \bibinfo{author}{\bibfnamefont{N.}~\bibnamefont{Djuric}},
  \bibinfo{author}{\bibfnamefont{G.~H.} \bibnamefont{Dunn}},
  \bibinfo{author}{\bibfnamefont{J.}~\bibnamefont{Semaniak}},
  \bibinfo{author}{\bibfnamefont{O.}~\bibnamefont{Novotny}},
  \bibinfo{author}{\bibfnamefont{A.}~\bibnamefont{Al-Khalili}},
  \bibinfo{author}{\bibfnamefont{A.}~\bibnamefont{Ehlerding}},
  \bibnamefont{et~al.}, \bibinfo{journal}{Nature (London)}
  \textbf{\bibinfo{volume}{422}}, \bibinfo{pages}{500} (\bibinfo{year}{2003}).

\bibitem[{\citenamefont{Kreckel et~al.}(2005)\citenamefont{Kreckel, Motsch, and
  \textit{et al.}}}]{tsr}
\bibinfo{author}{\bibfnamefont{H.}~\bibnamefont{Kreckel}},
  \bibinfo{author}{\bibfnamefont{M.}~\bibnamefont{Motsch}}, \bibnamefont{and}
  \bibinfo{author}{\bibfnamefont{J.~M.} \bibnamefont{\textit{et al.}}},
  \bibinfo{journal}{Phys. Rev. Lett.} \textbf{\bibinfo{volume}{95}},
  \bibinfo{pages}{263201} (\bibinfo{year}{2005}).

\bibitem[{\citenamefont{Curik and Greene}(2007{\natexlab{a}})}]{roman}
\bibinfo{author}{\bibfnamefont{R.}~\bibnamefont{Curik}} \bibnamefont{and}
  \bibinfo{author}{\bibfnamefont{C.~H.} \bibnamefont{Greene}},
  \bibinfo{journal}{Phys. Rev. Lett.} \textbf{\bibinfo{volume}{98}},
  \bibinfo{pages}{173201} (\bibinfo{year}{2007}{\natexlab{a}}).

\bibitem[{\citenamefont{Curik and Greene}(2007{\natexlab{b}})}]{roman2}
\bibinfo{author}{\bibfnamefont{R.}~\bibnamefont{Curik}} \bibnamefont{and}
  \bibinfo{author}{\bibfnamefont{C.~H.} \bibnamefont{Greene}},
  \bibinfo{journal}{Mol. Phys.} \textbf{\bibinfo{volume}{105}},
  \bibinfo{pages}{1565} (\bibinfo{year}{2007}{\natexlab{b}}).

\bibitem[{\citenamefont{Krohn}(2001)}]{krohn}
\bibinfo{author}{\bibfnamefont{S.}~\bibnamefont{Krohn}},
  \bibinfo{journal}{Ph.D. thesis, the University of Heidelberg, Germany}
  (\bibinfo{year}{2001}).

\bibitem[{\citenamefont{Krohn et~al.}(2001)\citenamefont{Krohn, Lange, Grieser,
  Knoll, Kreckel, Levin2, Repnow2, Schwalm2, Wester2, Witte et~al.}}]{krohnprl}
\bibinfo{author}{\bibfnamefont{S.}~\bibnamefont{Krohn}},
  \bibinfo{author}{\bibfnamefont{M.}~\bibnamefont{Lange}},
  \bibinfo{author}{\bibfnamefont{M.}~\bibnamefont{Grieser}},
  \bibinfo{author}{\bibfnamefont{L.}~\bibnamefont{Knoll}},
  \bibinfo{author}{\bibfnamefont{H.}~\bibnamefont{Kreckel}},
  \bibinfo{author}{\bibfnamefont{J.}~\bibnamefont{Levin2}},
  \bibinfo{author}{\bibfnamefont{R.}~\bibnamefont{Repnow2}},
  \bibinfo{author}{\bibfnamefont{D.}~\bibnamefont{Schwalm2}},
  \bibinfo{author}{\bibfnamefont{R.}~\bibnamefont{Wester2}},
  \bibinfo{author}{\bibfnamefont{P.}~\bibnamefont{Witte}},
  \bibnamefont{et~al.}, \bibinfo{journal}{Phys. Rev. Lett.}
  \textbf{\bibinfo{volume}{86}}, \bibinfo{pages}{4005} (\bibinfo{year}{2001}).

\bibitem[{\citenamefont{Tennyson and Morgan}(1999)}]{ukrmatrix}
\bibinfo{author}{\bibfnamefont{J.}~\bibnamefont{Tennyson}} \bibnamefont{and}
  \bibinfo{author}{\bibfnamefont{L.~A.} \bibnamefont{Morgan}},
  \bibinfo{journal}{Phil. Trans. R. Soc. Lond A}
  \textbf{\bibinfo{volume}{357}}, \bibinfo{pages}{1161} (\bibinfo{year}{1999}).

\bibitem[{\citenamefont{McCurdy et~al.}(2004)\citenamefont{McCurdy, Baertschy,
  and Rescigno}}]{ecs}
\bibinfo{author}{\bibfnamefont{C.~W.} \bibnamefont{McCurdy}},
  \bibinfo{author}{\bibfnamefont{M.}~\bibnamefont{Baertschy}},
  \bibnamefont{and} \bibinfo{author}{\bibfnamefont{T.~N.}
  \bibnamefont{Rescigno}}, \bibinfo{journal}{J. Phys. B}
  \textbf{\bibinfo{volume}{37}}, \bibinfo{pages}{R137} (\bibinfo{year}{2004}).

\bibitem[{\citenamefont{Lester}(1970)}]{lester1}
\bibinfo{author}{\bibfnamefont{W.~A.} \bibnamefont{Lester}},
  \bibinfo{journal}{J. Chem. Phys.} \textbf{\bibinfo{volume}{53}},
  \bibinfo{pages}{1151} (\bibinfo{year}{1970}).

\bibitem[{\citenamefont{Lester}(1971)}]{lester2}
\bibinfo{author}{\bibfnamefont{W.~A.} \bibnamefont{Lester}},
  \bibinfo{journal}{J. Chem. Phys.} \textbf{\bibinfo{volume}{54}},
  \bibinfo{pages}{3171} (\bibinfo{year}{1971}).

\bibitem[{\citenamefont{Kutzelnigg et~al.}(1973)\citenamefont{Kutzelnigg,
  Staemmler, and Hoheisel}}]{ksh}
\bibinfo{author}{\bibfnamefont{W.}~\bibnamefont{Kutzelnigg}},
  \bibinfo{author}{\bibfnamefont{V.}~\bibnamefont{Staemmler}},
  \bibnamefont{and} \bibinfo{author}{\bibfnamefont{C.}~\bibnamefont{Hoheisel}},
  \bibinfo{journal}{Chem. Phys.} \textbf{\bibinfo{volume}{1}},
  \bibinfo{pages}{27} (\bibinfo{year}{1973}).

\bibitem[{\citenamefont{Wagner and Wahl}(1978)}]{lih21978}
\bibinfo{author}{\bibfnamefont{A.~F.} \bibnamefont{Wagner}} \bibnamefont{and}
  \bibinfo{author}{\bibfnamefont{A.~C.} \bibnamefont{Wahl}},
  \bibinfo{journal}{J. Chem. Phys.} \textbf{\bibinfo{volume}{69}},
  \bibinfo{pages}{3756} (\bibinfo{year}{1978}).

\bibitem[{\citenamefont{Wu}(1979)}]{wu}
\bibinfo{author}{\bibfnamefont{C.~H.} \bibnamefont{Wu}}, \bibinfo{journal}{J.
  Chem. Phys.} \textbf{\bibinfo{volume}{71}}, \bibinfo{pages}{783}
  (\bibinfo{year}{1979}).

\bibitem[{\citenamefont{Dixon et~al.}(1988)\citenamefont{Dixon, Gole, and
  Kormornicki}}]{dgk}
\bibinfo{author}{\bibfnamefont{D.~A.} \bibnamefont{Dixon}},
  \bibinfo{author}{\bibfnamefont{J.~L.} \bibnamefont{Gole}}, \bibnamefont{and}
  \bibinfo{author}{\bibfnamefont{A.}~\bibnamefont{Kormornicki}},
  \bibinfo{journal}{J. Phys. Chem.} \textbf{\bibinfo{volume}{92}},
  \bibinfo{pages}{1378} (\bibinfo{year}{1988}).

\bibitem[{\citenamefont{Searles and von Nagy-Felsobuki}(1991)}]{sn}
\bibinfo{author}{\bibfnamefont{D.~J.} \bibnamefont{Searles}} \bibnamefont{and}
  \bibinfo{author}{\bibfnamefont{E.~I.} \bibnamefont{von Nagy-Felsobuki}},
  \bibinfo{journal}{Phys. Rev. A} \textbf{\bibinfo{volume}{43}},
  \bibinfo{pages}{3365} (\bibinfo{year}{1991}).

\bibitem[{\citenamefont{Dunne et~al.}(2001)\citenamefont{Dunne, Murrell, and
  Jemmer}}]{dunne}
\bibinfo{author}{\bibfnamefont{L.~J.} \bibnamefont{Dunne}},
  \bibinfo{author}{\bibfnamefont{J.~N.} \bibnamefont{Murrell}},
  \bibnamefont{and} \bibinfo{author}{\bibfnamefont{P.}~\bibnamefont{Jemmer}},
  \bibinfo{journal}{Chem. Phys. Lett.} \textbf{\bibinfo{volume}{336}},
  \bibinfo{pages}{1} (\bibinfo{year}{2001}).

\bibitem[{\citenamefont{Martinazzo et~al.}(2003)\citenamefont{Martinazzo,
  Tantardini, Bodo, and Gianturco}}]{lih2surf}
\bibinfo{author}{\bibfnamefont{R.}~\bibnamefont{Martinazzo}},
  \bibinfo{author}{\bibfnamefont{G.~F.} \bibnamefont{Tantardini}},
  \bibinfo{author}{\bibfnamefont{E.}~\bibnamefont{Bodo}}, \bibnamefont{and}
  \bibinfo{author}{\bibfnamefont{F.~A.} \bibnamefont{Gianturco}},
  \bibinfo{journal}{J. Chem. Phys.} \textbf{\bibinfo{volume}{119}},
  \bibinfo{pages}{11241} (\bibinfo{year}{2003}).

\bibitem[{\citenamefont{Ema et~al.}(2003)\citenamefont{Ema, de~la Vega,
  Ramirez, Lopez, Rico, Meissner, and Paldus}}]{lih2basis}
\bibinfo{author}{\bibfnamefont{I.}~\bibnamefont{Ema}},
  \bibinfo{author}{\bibfnamefont{J.~M.~G.} \bibnamefont{de~la Vega}},
  \bibinfo{author}{\bibfnamefont{G.}~\bibnamefont{Ramirez}},
  \bibinfo{author}{\bibfnamefont{R.}~\bibnamefont{Lopez}},
  \bibinfo{author}{\bibfnamefont{J.~F.} \bibnamefont{Rico}},
  \bibinfo{author}{\bibfnamefont{H.}~\bibnamefont{Meissner}}, \bibnamefont{and}
  \bibinfo{author}{\bibfnamefont{J.}~\bibnamefont{Paldus}},
  \bibinfo{journal}{J. Comput. Chem.} \textbf{\bibinfo{volume}{24}},
  \bibinfo{pages}{859} (\bibinfo{year}{2003}).

\bibitem[{\citenamefont{Alexandre et~al.}(2002)\citenamefont{Alexandre,
  Gorfinkiel, Morgan, and Tennyson}}]{gtobas}
\bibinfo{author}{\bibfnamefont{F.}~\bibnamefont{Alexandre}},
  \bibinfo{author}{\bibfnamefont{J.~D.} \bibnamefont{Gorfinkiel}},
  \bibinfo{author}{\bibfnamefont{L.~A.} \bibnamefont{Morgan}},
  \bibnamefont{and} \bibinfo{author}{\bibfnamefont{J.}~\bibnamefont{Tennyson}},
  \bibinfo{journal}{Computer Physics Communications}
  \textbf{\bibinfo{volume}{144}}, \bibinfo{pages}{224} (\bibinfo{year}{2002}).

\bibitem[{\citenamefont{Sukiasyan and Meyer}(2001)}]{hd2}
\bibinfo{author}{\bibfnamefont{S.}~\bibnamefont{Sukiasyan}} \bibnamefont{and}
  \bibinfo{author}{\bibfnamefont{H.-D.} \bibnamefont{Meyer}},
  \bibinfo{journal}{J. Phys. Chem. A} \textbf{\bibinfo{volume}{105}},
  \bibinfo{pages}{2604} (\bibinfo{year}{2001}).

\bibitem[{\citenamefont{Macek}(1968)}]{macek1968}
\bibinfo{author}{\bibfnamefont{J.}~\bibnamefont{Macek}}, \bibinfo{journal}{J.
  Phys. B} \textbf{\bibinfo{volume}{1}}, \bibinfo{pages}{831}
  (\bibinfo{year}{1968}).

\bibitem[{\citenamefont{Delves}(1959)}]{delves1}
\bibinfo{author}{\bibfnamefont{L.~M.} \bibnamefont{Delves}},
  \bibinfo{journal}{Nucl. Phys.} \textbf{\bibinfo{volume}{9}},
  \bibinfo{pages}{391} (\bibinfo{year}{1959}).

\bibitem[{\citenamefont{Delves}(1960)}]{delves2}
\bibinfo{author}{\bibfnamefont{L.~M.} \bibnamefont{Delves}},
  \bibinfo{journal}{Nucl. Phys.} \textbf{\bibinfo{volume}{20}},
  \bibinfo{pages}{275} (\bibinfo{year}{1960}).

\bibitem[{\citenamefont{Tennyson and Sutcliffe}(1982)}]{rembed}
\bibinfo{author}{\bibfnamefont{J.}~\bibnamefont{Tennyson}} \bibnamefont{and}
  \bibinfo{author}{\bibfnamefont{B.~T.} \bibnamefont{Sutcliffe}},
  \bibinfo{journal}{J. Chem. Phys.} \textbf{\bibinfo{volume}{77}},
  \bibinfo{pages}{4061} (\bibinfo{year}{1982}).

\bibitem[{\citenamefont{Petrongolo}(1988)}]{petrongolo}
\bibinfo{author}{\bibfnamefont{C.}~\bibnamefont{Petrongolo}},
  \bibinfo{journal}{J. Chem. Phys.} \textbf{\bibinfo{volume}{89}},
  \bibinfo{pages}{1297} (\bibinfo{year}{1988}).

\bibitem[{\citenamefont{Dickinson and Certain}(1968)}]{dickcert}
\bibinfo{author}{\bibfnamefont{A.~S.} \bibnamefont{Dickinson}}
  \bibnamefont{and} \bibinfo{author}{\bibfnamefont{P.~R.}
  \bibnamefont{Certain}}, \bibinfo{journal}{J Chem Phys}
  \textbf{\bibinfo{volume}{49}}, \bibinfo{pages}{4209} (\bibinfo{year}{1968}).

\bibitem[{\citenamefont{Light et~al.}(1985)\citenamefont{Light, Hamilton, and
  Lill}}]{lhl}
\bibinfo{author}{\bibfnamefont{J.~C.} \bibnamefont{Light}},
  \bibinfo{author}{\bibfnamefont{I.~P.} \bibnamefont{Hamilton}},
  \bibnamefont{and} \bibinfo{author}{\bibfnamefont{J.~V.} \bibnamefont{Lill}},
  \bibinfo{journal}{J Chem Phys} \textbf{\bibinfo{volume}{82}},
  \bibinfo{pages}{1400} (\bibinfo{year}{1985}).

\bibitem[{\citenamefont{Haxton}(2007)}]{lebdvr}
\bibinfo{author}{\bibfnamefont{D.~J.} \bibnamefont{Haxton}},
  \bibinfo{journal}{J. Phys. B} \textbf{\bibinfo{volume}{40}},
  \bibinfo{pages}{4443} (\bibinfo{year}{2007}).

\bibitem[{\citenamefont{Rescigno and McCurdy}(2000)}]{femdvr}
\bibinfo{author}{\bibfnamefont{T.~N.} \bibnamefont{Rescigno}} \bibnamefont{and}
  \bibinfo{author}{\bibfnamefont{C.~W.} \bibnamefont{McCurdy}},
  \bibinfo{journal}{Phys. Rev. A} \textbf{\bibinfo{volume}{62}},
  \bibinfo{pages}{032706} (\bibinfo{year}{2000}).

\bibitem[{\citenamefont{Sanz et~al.}(2005)\citenamefont{Sanz, Bodo, and
  Gianturco}}]{gianvib}
\bibinfo{author}{\bibfnamefont{C.}~\bibnamefont{Sanz}},
  \bibinfo{author}{\bibfnamefont{E.}~\bibnamefont{Bodo}}, \bibnamefont{and}
  \bibinfo{author}{\bibfnamefont{F.~A.} \bibnamefont{Gianturco}},
  \bibinfo{journal}{Chem. Phys.} \textbf{\bibinfo{volume}{314}},
  \bibinfo{pages}{135} (\bibinfo{year}{2005}).

\bibitem[{\citenamefont{Tolstikhin et~al.}(1996)\citenamefont{Tolstikhin,
  Watanabe, and Matsuzawa}}]{svd}
\bibinfo{author}{\bibfnamefont{O.~I.} \bibnamefont{Tolstikhin}},
  \bibinfo{author}{\bibfnamefont{S.}~\bibnamefont{Watanabe}}, \bibnamefont{and}
  \bibinfo{author}{\bibfnamefont{M.}~\bibnamefont{Matsuzawa}},
  \bibinfo{journal}{J. Phys. B} \textbf{\bibinfo{volume}{29}},
  \bibinfo{pages}{L389} (\bibinfo{year}{1996}).

\bibitem[{\citenamefont{Tolstikhin et~al.}(1998)\citenamefont{Tolstikhin,
  Ostrovsky, and Nakamura}}]{pseudo}
\bibinfo{author}{\bibfnamefont{O.~I.} \bibnamefont{Tolstikhin}},
  \bibinfo{author}{\bibfnamefont{V.~N.} \bibnamefont{Ostrovsky}},
  \bibnamefont{and} \bibinfo{author}{\bibfnamefont{H.}~\bibnamefont{Nakamura}},
  \bibinfo{journal}{Phys. Rev. A} \textbf{\bibinfo{volume}{58}},
  \bibinfo{pages}{2077} (\bibinfo{year}{1998}).

\bibitem[{\citenamefont{Jackle and Meyer}(1996)}]{CAPform}
\bibinfo{author}{\bibfnamefont{A.}~\bibnamefont{Jackle}} \bibnamefont{and}
  \bibinfo{author}{\bibfnamefont{H.-D.} \bibnamefont{Meyer}},
  \bibinfo{journal}{J. Chem. Phys.} \textbf{\bibinfo{volume}{105}},
  \bibinfo{pages}{6778} (\bibinfo{year}{1996}).

\bibitem[{\citenamefont{Leforestier and Wyatt}(1983)}]{CAPref1}
\bibinfo{author}{\bibfnamefont{C.}~\bibnamefont{Leforestier}} \bibnamefont{and}
  \bibinfo{author}{\bibfnamefont{R.~E.} \bibnamefont{Wyatt}},
  \bibinfo{journal}{J. Chem. Phys} \textbf{\bibinfo{volume}{78}}
  (\bibinfo{year}{1983}).

\bibitem[{\citenamefont{Kosloff and Kosloff}(1986)}]{CAPref2}
\bibinfo{author}{\bibfnamefont{R.}~\bibnamefont{Kosloff}} \bibnamefont{and}
  \bibinfo{author}{\bibfnamefont{D.}~\bibnamefont{Kosloff}},
  \bibinfo{journal}{J. Comput. Phys.} \textbf{\bibinfo{volume}{63}}
  (\bibinfo{year}{1986}).

\bibitem[{\citenamefont{Aguilar and Combes}(1971)}]{abc1}
\bibinfo{author}{\bibfnamefont{J.}~\bibnamefont{Aguilar}} \bibnamefont{and}
  \bibinfo{author}{\bibfnamefont{J.~M.} \bibnamefont{Combes}},
  \bibinfo{journal}{Commun. Math. Phys.} \textbf{\bibinfo{volume}{22}},
  \bibinfo{pages}{269} (\bibinfo{year}{1971}).

\bibitem[{\citenamefont{Balslev and Combes}(1971)}]{abc2}
\bibinfo{author}{\bibfnamefont{E.}~\bibnamefont{Balslev}} \bibnamefont{and}
  \bibinfo{author}{\bibfnamefont{J.~M.} \bibnamefont{Combes}},
  \bibinfo{journal}{Commun. Math. Phys.} \textbf{\bibinfo{volume}{22}},
  \bibinfo{pages}{280} (\bibinfo{year}{1971}).

\bibitem[{\citenamefont{Moiseyev et~al.}(1978)\citenamefont{Moiseyev, Certain,
  and Weinhold}}]{moi2}
\bibinfo{author}{\bibfnamefont{N.}~\bibnamefont{Moiseyev}},
  \bibinfo{author}{\bibfnamefont{P.~R.} \bibnamefont{Certain}},
  \bibnamefont{and} \bibinfo{author}{\bibfnamefont{F.}~\bibnamefont{Weinhold}},
  \bibinfo{journal}{Mol. Phys.} \textbf{\bibinfo{volume}{36}},
  \bibinfo{pages}{1613} (\bibinfo{year}{1978}).

\bibitem[{\citenamefont{Moiseyev and Hirschfelder}(1987)}]{moi3}
\bibinfo{author}{\bibfnamefont{N.}~\bibnamefont{Moiseyev}} \bibnamefont{and}
  \bibinfo{author}{\bibfnamefont{J.~O.} \bibnamefont{Hirschfelder}},
  \bibinfo{journal}{J. Chem. Phys.} \textbf{\bibinfo{volume}{88}},
  \bibinfo{pages}{1063} (\bibinfo{year}{1987}).

\bibitem[{\citenamefont{Lipkin et~al.}(1992)\citenamefont{Lipkin, Lefebvre, and
  Moiseyev}}]{moi4}
\bibinfo{author}{\bibfnamefont{N.}~\bibnamefont{Lipkin}},
  \bibinfo{author}{\bibfnamefont{R.}~\bibnamefont{Lefebvre}}, \bibnamefont{and}
  \bibinfo{author}{\bibfnamefont{N.}~\bibnamefont{Moiseyev}},
  \bibinfo{journal}{Phys. Rev. A} \textbf{\bibinfo{volume}{45}},
  \bibinfo{pages}{4553} (\bibinfo{year}{1992}).

\bibitem[{\citenamefont{Moiseyev}(1998)}]{moirev}
\bibinfo{author}{\bibfnamefont{N.}~\bibnamefont{Moiseyev}},
  \bibinfo{journal}{Physics Reports} \textbf{\bibinfo{volume}{302}},
  \bibinfo{pages}{211} (\bibinfo{year}{1998}).

\bibitem[{\citenamefont{Fano}(1970)}]{fanoft}
\bibinfo{author}{\bibfnamefont{U.}~\bibnamefont{Fano}}, \bibinfo{journal}{Phys.
  Rev. A} \textbf{\bibinfo{volume}{2}}, \bibinfo{pages}{353}
  (\bibinfo{year}{1970}).

\bibitem[{\citenamefont{Chang and Fano}(1972)}]{changfano}
\bibinfo{author}{\bibfnamefont{E.~S.} \bibnamefont{Chang}} \bibnamefont{and}
  \bibinfo{author}{\bibfnamefont{U.}~\bibnamefont{Fano}},
  \bibinfo{journal}{Phys. Rev. A} \textbf{\bibinfo{volume}{6}},
  \bibinfo{pages}{173} (\bibinfo{year}{1972}).

\bibitem[{\citenamefont{Armstrong and Greene}(1994)}]{starkft}
\bibinfo{author}{\bibfnamefont{D.~J.} \bibnamefont{Armstrong}}
  \bibnamefont{and} \bibinfo{author}{\bibfnamefont{C.~H.}
  \bibnamefont{Greene}}, \bibinfo{journal}{Phys. Rev. A}
  \textbf{\bibinfo{volume}{50}}, \bibinfo{pages}{4956} (\bibinfo{year}{1994}).

\bibitem[{\citenamefont{Lee and Lu}(1973)}]{lsjj0}
\bibinfo{author}{\bibfnamefont{C.~M.} \bibnamefont{Lee}} \bibnamefont{and}
  \bibinfo{author}{\bibfnamefont{K.~T.} \bibnamefont{Lu}},
  \bibinfo{journal}{Phys. Rev. A} \textbf{\bibinfo{volume}{8}},
  \bibinfo{pages}{1241} (\bibinfo{year}{1973}).

\bibitem[{\citenamefont{Robicheaux and Greene}(1993)}]{lsjj}
\bibinfo{author}{\bibfnamefont{F.}~\bibnamefont{Robicheaux}} \bibnamefont{and}
  \bibinfo{author}{\bibfnamefont{C.~H.} \bibnamefont{Greene}},
  \bibinfo{journal}{Phys. Rev. A} \textbf{\bibinfo{volume}{47}},
  \bibinfo{pages}{4908} (\bibinfo{year}{1993}).

\bibitem[{\citenamefont{Jungen and Raseev}(1998)}]{jungenraseev}
\bibinfo{author}{\bibfnamefont{C.}~\bibnamefont{Jungen}} \bibnamefont{and}
  \bibinfo{author}{\bibfnamefont{G.}~\bibnamefont{Raseev}},
  \bibinfo{journal}{Phys. Rev. A} \textbf{\bibinfo{volume}{57}},
  \bibinfo{pages}{2407} (\bibinfo{year}{1998}).

\bibitem[{\citenamefont{Chase}(1956)}]{chase}
\bibinfo{author}{\bibfnamefont{D.~M.} \bibnamefont{Chase}},
  \bibinfo{journal}{Phys. Rev. A} \textbf{\bibinfo{volume}{104}},
  \bibinfo{pages}{838} (\bibinfo{year}{1956}).

\bibitem[{\citenamefont{Seaton}(1983)}]{seaton}
\bibinfo{author}{\bibfnamefont{M.~J.} \bibnamefont{Seaton}},
  \bibinfo{journal}{Rep. Prog. Phys.} \textbf{\bibinfo{volume}{46}},
  \bibinfo{pages}{167} (\bibinfo{year}{1983}).

\bibitem[{\citenamefont{Greene et~al.}(1979)\citenamefont{Greene, Fano, and
  Strinati}}]{general}
\bibinfo{author}{\bibfnamefont{C.}~\bibnamefont{Greene}},
  \bibinfo{author}{\bibfnamefont{U.}~\bibnamefont{Fano}}, \bibnamefont{and}
  \bibinfo{author}{\bibfnamefont{G.}~\bibnamefont{Strinati}},
  \bibinfo{journal}{Phys. Rev. A} \textbf{\bibinfo{volume}{19}},
  \bibinfo{pages}{1485} (\bibinfo{year}{1979}).

\bibitem[{\citenamefont{Greene et~al.}(1982)\citenamefont{Greene, Rau, and
  Fano}}]{general2}
\bibinfo{author}{\bibfnamefont{C.~H.} \bibnamefont{Greene}},
  \bibinfo{author}{\bibfnamefont{A.~R.~P.} \bibnamefont{Rau}},
  \bibnamefont{and} \bibinfo{author}{\bibfnamefont{U.}~\bibnamefont{Fano}},
  \bibinfo{journal}{Phys. Rev. A} \textbf{\bibinfo{volume}{26}},
  \bibinfo{pages}{2441} (\bibinfo{year}{1982}).

\bibitem[{\citenamefont{Ross and Jungen}(1994{\natexlab{a}})}]{rossh2_1}
\bibinfo{author}{\bibfnamefont{S.~C.} \bibnamefont{Ross}} \bibnamefont{and}
  \bibinfo{author}{\bibfnamefont{C.}~\bibnamefont{Jungen}},
  \bibinfo{journal}{Phys. Rev. A} \textbf{\bibinfo{volume}{49}},
  \bibinfo{pages}{4353} (\bibinfo{year}{1994}{\natexlab{a}}).

\bibitem[{\citenamefont{Ross and Jungen}(1994{\natexlab{b}})}]{rossh2_2}
\bibinfo{author}{\bibfnamefont{S.~C.} \bibnamefont{Ross}} \bibnamefont{and}
  \bibinfo{author}{\bibfnamefont{C.}~\bibnamefont{Jungen}},
  \bibinfo{journal}{Phys. Rev. A} \textbf{\bibinfo{volume}{49}},
  \bibinfo{pages}{4364} (\bibinfo{year}{1994}{\natexlab{b}}).

\bibitem[{\citenamefont{Gao and Greene}(1989)}]{gaogreene}
\bibinfo{author}{\bibfnamefont{H.}~\bibnamefont{Gao}} \bibnamefont{and}
  \bibinfo{author}{\bibfnamefont{C.~H.} \bibnamefont{Greene}},
  \bibinfo{journal}{J. Chem. Phys} \textbf{\bibinfo{volume}{91}},
  \bibinfo{pages}{3988} (\bibinfo{year}{1989}).

\bibitem[{\citenamefont{Gao and Greene}(1990)}]{gaogreene2}
\bibinfo{author}{\bibfnamefont{H.}~\bibnamefont{Gao}} \bibnamefont{and}
  \bibinfo{author}{\bibfnamefont{C.~H.} \bibnamefont{Greene}},
  \bibinfo{journal}{Phys. Rev. A} \textbf{\bibinfo{volume}{42}},
  \bibinfo{pages}{6946} (\bibinfo{year}{1990}).

\bibitem[{\citenamefont{Gao}(1992)}]{gaoh2}
\bibinfo{author}{\bibfnamefont{H.}~\bibnamefont{Gao}}, \bibinfo{journal}{Phys.
  Rev. A} \textbf{\bibinfo{volume}{45}}, \bibinfo{pages}{6895}
  (\bibinfo{year}{1992}).

\bibitem[{\citenamefont{Jungen}(1984)}]{unified}
\bibinfo{author}{\bibfnamefont{C.}~\bibnamefont{Jungen}},
  \bibinfo{journal}{Phys. Rev. Lett.} \textbf{\bibinfo{volume}{53}},
  \bibinfo{pages}{2394} (\bibinfo{year}{1984}).

\bibitem[{\citenamefont{Lee}(1977)}]{leemqdt}
\bibinfo{author}{\bibfnamefont{C.~M.} \bibnamefont{Lee}},
  \bibinfo{journal}{Phys. Rev. A} \textbf{\bibinfo{volume}{16}},
  \bibinfo{pages}{109} (\bibinfo{year}{1977}).

\bibitem[{\citenamefont{Child and Jungen}(1990)}]{childjungen}
\bibinfo{author}{\bibfnamefont{M.~S.} \bibnamefont{Child}} \bibnamefont{and}
  \bibinfo{author}{\bibfnamefont{C.}~\bibnamefont{Jungen}},
  \bibinfo{journal}{J. Chem. Phys.} \textbf{\bibinfo{volume}{93}},
  \bibinfo{pages}{7756} (\bibinfo{year}{1990}).

\end{thebibliography}

\end{document}